\documentclass[prd,preprint,superscriptaddress,nofootinbib,%
tightenlines]{revtex4}
\usepackage{mathrsfs}
\usepackage{latexsym,bm}
\usepackage{graphicx}
\usepackage{indentfirst}
\usepackage{slashed}
\usepackage{amssymb}
\usepackage{amsmath}
\usepackage{bbm}
\usepackage{color}
\usepackage[dvipsnames]{xcolor}
\usepackage{epsfig}
\usepackage[titletoc]{appendix}
\usepackage{multirow}%
\usepackage{rotating}
\usepackage{epstopdf}
\usepackage{extarrows}
\usepackage{tabularx}
\usepackage{soul} 
\usepackage{xcolor}
\usepackage{lipsum}

\usepackage{caption}
\usepackage{subcaption}
\usepackage[colorlinks=true,allcolors=BlueViolet,hyperfootnotes=false]{hyperref}

\usepackage[utf8]{inputenc}
\usepackage{slashed}

\newcommand{\be}{\begin{equation}} \newcommand{\ee}{\end{equation}}
\newcommand{\ba}{\begin{array}{c}} \newcommand{\ea}{\end{array}}
\newcommand{\bea}{\begin{eqnarray}} \newcommand{\eea}{\end{eqnarray}}

\begin{document}
\title{Weak decays of $B_s$ to $D_s$ based on the helicity analysis}
\author{Sara Rahmani}
\email{s.rahmani120@gmail.com}
\affiliation{School of Physics and Electronics, Hunan University, Changsha 410082, China}

\begin{abstract}
Employing dipole and exponential hadronic transition form factors and helicity analysis combined with the Lattice QCD input, we present a detailed study of the decays $B_s \to D_s^{(*)}\ell \nu_\ell$. $B_s \to D_s + P(V)$, where $P = \pi^+, K^+, D_s^+$ and $V = \rho^+, K^{*+}, D_s^{*+}$, are also investigated and their branching fractions are examined.
At the large recoil point, we calculate $f_{ + ,0}^{{B_s} \to {D_s}}(0) = 0.67 \pm 0.01$ for both parameterizations. Then we evaluate the branching fractions $BR({B_s} \to {D_s}\ell \nu )$ and $BR({B_s} \to {D_s}\ell \nu )$, which leads to $R({D_s}) = 0.298 \pm 0.123$. The ratios $\frac{{\Gamma (B_s^0 \to D_s^ - {\mu ^ + }{\nu _\mu })}}{{\Gamma (B_s^0 \to D_s^{* - }{\mu ^ + }{\nu _\mu })}}$ are found to be $0.4415 \pm 0.1860$ and $0.4406 \pm 0.1854$, which are in good agreement of recent LHCb collaboration measurement.
We also calculate the physical observables, $A_{FB}^\ell ,P_L^\ell ,P_T^\ell ,C_F^\ell ,F_L^\ell $.
\end{abstract}

\maketitle

\section{Introduction}
The weak decay processes of $B_s$ have been one of the most fascinating subjects for testing the Standard Model (SM) and provide a good source of knowledge in physics beyond SM, considering that the lepton flavor universality can be explored in these types of decay processes \cite{Gershtein:1976aq,Khlopov:1978id}. All electroweak gauge bosons, $Z^0, \gamma, W^{\pm}$ have equivalent interactions to the three leptonic generations in lepton flavor universality and the only divergence comes from the mass distinctions of $e, \mu, \tau$.

The experimental predictions of some physical observables such as ratios ${R_D},{R_{{D^*}}},{R_{J/\psi }}$ and longitudinal polarization of the final vector meson $D^*$, $F_L^\tau $ and the longitudinal polarization of the lepton $\tau$, $P_L^\tau $ in $B$ decays mediated by the quark level $b \to c\ell {{\bar \nu }_\ell }$ have disparities from SM predictions.
In 2020, LHCb collaboration reported the first measurements of $B_s^0 \to D_s^{*-} {\mu ^ + }{\nu _\mu }$ to be $BR(B_s^0 \to D_s^ - {\mu ^ + }{\nu _\mu }) = (2.49 \pm 0.12 \pm 0.14 \pm 0.16)$ and $BR(B_s^0 \to D_s^{* - }{\mu ^ + }{\nu _\mu }) = (5.38 \pm 0.25 \pm 0.46 \pm 0.30)$ \cite{LHCb:2020cyw}, where the first, second and third uncertainties refer to statistical, systematic, and from the external inputs respectively. They also determined the ratio of $BR(B_s^0 \to D_s^ - {\mu ^ + }{\nu _\mu })/BR(B_s^0 \to D_s^{* - }{\mu ^ + }{\nu _\mu })$ to be $0.464 \pm 0.013 \pm 0.043$. Belle Collaboration reported the ratios of branching fractions $BR(\bar B \to {D^{(*)}}{\tau ^ - }{{\bar \nu }_\tau })/BR(\bar B \to {D^{(*)}}{\ell ^ - }{{\bar \nu }_\ell })$ as $R(D) = 0.307 \pm 0.037 \pm 0.016$ and $R({D^*}) = 0.283 \pm 0.018 \pm 0.014$ in $B$ decays measurements \cite{Belle:2019rba}. The SM predictions are $0.298 \pm 0.004$ for $R(D)$ and $0.254 \pm 0.005$ in case of $R(D^*)$ in Ref. \cite{HFLAV:2022esi}. Heavy Flavor Averaging Group Collaboration also reported the branching fractions of two body decays of $B_s^0$ as follows: $BR(B_s^0 \to D_s^ - {\pi ^ + }) = (2.85 \pm 0.18) \times {10^{ - 3}}$, $BR(B_s^0 \to D_s^ - {K^ + }) = (0.213 \pm 0.014) \times {10^{ - 3}}$, and $BR(B_s^0 \to D_s^ - {\rho ^ + }) = (7.7_{ - 1.8}^{ + 1.9}) \times {10^{ - 3}}$ \cite{HFLAV:2022esi}.
The transition form factors of $B_s \to D_s$ have been studied by using the light-cone sum rules as well as heavy quark effective field theory and discussed the branching fractions of semileptonic $BR({B_s} \to {D_s}\ell \nu ), \ell = e, \mu, \tau$ \cite{Zhang:2022opp}. Hu \textit{et al.} studied semileptonic $B$ decays using perturbative QCD factorization with the Lattice QCD input \cite{Hu:2019bdf}.
Semielptonic and noleptonic decays of $B_s$ have been investigated in the nonrelativistic constituent quark models \cite{Albertus:2014eqa}. The bottom transitions to the charm ones are discussed within Isgur-Wise functions in Refs. \cite{Rahmani:2022vix, Xiao:2020gry, Rahmani:2017vbg, Rahmani:2017exi}.
Zhou \textit{et al.} evaluated the ratios of $R(D_s^{(*)}), R(D^{(*)})$ based on improved instantaneous Bethe–Salpeter method and relativistic corrections of the form factors \cite{Zhou:2019stx}. Cui \textit{et al.} computed the next-to-leading-order QCD 
computations of the $B_s \to D_s^{(*)} \ell \nu$ \cite{Cui:2023jiw}. 
Angular analysis of $b\to c$ decays have been performed and several observables in SM and beyond have been studied \cite{Rajeev:2018txm,Das:2021lws}.
Lattice QCD determinations of the decays $B_s \to D_s \ell \nu$ and $B_s \to D_s^* \ell \nu$ have been done by HPQCD Collaborations in Ref. \cite{McLean:2019qcx} and Ref. \cite{Harrison:2021tol} respectively. We will use their data to perform fits in our model.
\par
Motivated by the tensions between experimental measurements and theoretical SM predictions, we aim to study modes involving $B_s \to D_s^{(*)}$ in both semileptonic and nonleptoinc sectors. In the next section, we will present our formalism including helicity amplitudes \ref{section: helicity}, twofold angular distribution \ref{section: decay width}, physical observables \ref{sec: observables}, the formulations of two body decays $B_s \to D_sP$ and $B_s \to D_sV$ \ref{sec: twobody}, and form factors \ref{sec: form factors}. In section \ref{section: numeric}, we will give a detailed numerical analysis and discuss the results of section \ref{section: formalism}. Finally, Concluding remarks are provided in section \ref{section: conclusion}.

\section{Formalism}
\label{section: formalism}
\subsection{Helicity amplitudes}
\label{section: helicity}
The effective Fermi Lagrangian for the semileptonic $b$ to $c$ induced transitions reads as
\begin{equation}
    \mathcal{L}_{eff} = \frac{{{G_F}}}{{\sqrt 2 }}{V_{cb}}\bar \ell{\gamma ^\mu }(1 - {\gamma _5}){\nu _\ell}\bar c{\gamma _\mu }(1 - {\gamma _5})b,
\end{equation}
in which, 
\begin{equation}
    \begin{array}{l}
{\cal M}{\rm{ = }}\frac{{{G_F}}}{{\sqrt 2 }}{V_{cb}}\left\langle {{D_s^{(*)}}|\bar c{{\cal O}^\mu }b|{B_s}} \right\rangle {\ell^ + }{{\cal O}_\mu }{\nu _\ell}\\
{\rm{ = }}\frac{{{G_F}}}{{\sqrt 2 }}{V_{cb}}{H^\mu }{L_\mu },
\end{array}
\end{equation}
where ${{\cal O}^\mu } = {\gamma ^\mu }(1 - {\gamma _5})$, and $G_F$ is the Fermi coupling constant. In the electroweak current $ < {D_s}|{(V - A)_\mu }|{B_s} > $, the vector component is ${V_\mu } = \bar b{\gamma _\mu }c$ and ${A_\mu } = \bar b{\gamma_5\gamma _\mu }c$ is the axial vector one. In the amplitude of $B_s \to D_s$, only the vector component contributes, because axial vector one fails to meet the property of parity invariance of QCD. The transition of $B_s \to D_s \ell^+\nu _\ell$ in terms of form factors can be written as
\begin{equation}
\label{eq: pseu form}
     < {D_s}({p_2})|{V_\mu }|{B_s}({p_1}) >  = ({P_\mu } - \frac{{m_1^2 - m_2^2}}{{{q^2}}}{q_\mu }){F_1}({q^2}) + \frac{{m_1^2 - m_2^2}}{{{q^2}}}{q_\mu }{F_0}({q^2}),
\end{equation}
where $p_1$, $p_2$, $m_1$, and $m_2$ are the four-momentum of $B_s$, $D_s$, and their masses respectively. $P_\mu$ and $q_\mu$ are explained as ${P_\mu } = {({p_1} + {p_2})_\mu }$ and ${q_\mu } = {({q_1} + {q_2})_\mu }$.
For the transition of $B_s \to D_s^* \ell^+\nu _\ell$, one can write
\begin{equation}
     < D_s^*({p_2})|{V_\mu } - {A_\mu }|{B_s}({p_1}) >  = \epsilon _2^{{\text{\dag }}\nu }{T_{\mu \nu }},
\end{equation}
with $\epsilon_2$ the polarization vector of the vector meson $D_s^*$ and one has
\begin{equation}
    \begin{gathered}
    \label{eq: vector amp}
  {T_{\mu \nu }} =  - ({m_1} + {m_2})[{g_{\mu \nu }} - \frac{{{P_\nu }}}{{{q^2}}}{q_\mu }]{A_1}({q^2}) + \frac{{{P_\nu }}}{{{m_1} + {m_2}}}[{P_\mu } - \frac{{m_1^2 - m_2^2}}{{{q^2}}}{q_\mu }]{A_2}({q^2}) \hfill \\
   - 2{m_2}\frac{{{P_\nu }}}{{{q^2}}}{q_\mu }{A_0}({q^2}) + \frac{i}{{{m_1} + {m_2}}}{\varepsilon _{\mu \nu \alpha \beta }}{P^\alpha }{q^\beta }V({q^2}). \hfill \\ 
\end{gathered} 
\end{equation}
The Levi-Civita tensor in Minkowski space is ${\varepsilon _{0123}} =  + 1$. For $B_s \to D_s^*$, the hadronic tensor can be written as \cite{Zhang:2020dla}
\begin{equation}
    {H^{\mu \nu }} = {T_{\mu \alpha }}T_{\nu \beta }^{\text{\dag }}( - {g^{\alpha \beta }} + \frac{{p_2^\alpha p_2^\beta }}{{m_2^2}}),
\end{equation}
in which we consider ${H_\mu } = \varepsilon _2^{{\text{*}}\alpha }{T_{\mu \alpha }}$, $|p_2|$ the momentum of the $W_\text{off-shell}$ and the relation between polarization vectors $\epsilon^\mu \epsilon ^{\text{*}\nu}  =  - {g^{\mu \nu }} + \frac{{p_2^\mu p_2^\nu }}{{m_2^2}}$.
\par
In the helicity space, the hadronic tensor is given by
\begin{equation}
\label{eq: hadronic tens pseu}
    H({\lambda _W},{{\lambda '}_W}) = \epsilon^{\dag \mu} (\lambda_W) \epsilon^\nu({\lambda '}_W)H_{\mu\nu}=H_{\lambda_W}H^\dag_{\lambda '_W},
\end{equation}
for $B_s \to D_s$ using $H_{\lambda_W} = \epsilon^{\dag \mu} (\lambda_W)T_\mu$ in which $\lambda_W$ describes the polarization index of $W_\text{off-shell}$. In the helicity basis of $\epsilon^\mu(\lambda_W)$, there are an orthonormal and complete polarization vectors with three spin-1 components having the relation with momentum transfer $\epsilon^\mu({\lambda _W}){q_\mu } = 0$ in the cases of ${\lambda _W} = 0, \pm$, and $\epsilon^\mu({t}) = {q_\mu }/ \sqrt {{q^2}} $ with one spin-0 component ${\lambda _W} = t$. The polarization four-vectors have orthonormality and completeness relations
\begin{equation}
\label{Eq: ortho}
    \epsilon_\mu(\lambda_W)\epsilon^\dag_\nu(\lambda '_W)g_{\lambda_W\lambda'_W}=g_{\mu\nu},
\end{equation}
and
\begin{equation}
\label{Eq: complete}
    \epsilon_\mu^\dag(\lambda_W)\epsilon^\mu(\lambda '_W)=g_{\lambda_W\lambda'_W},
\end{equation}
with $g_{\mu\nu} = \text{diag}(+, -, -, -)$. Using Eqs. (\ref{Eq: ortho}, \ref{Eq: complete}) in the helicity component space, the leptonic and hadronic tensors will be contracted as \cite{Zhang:2020dla,Ivanov:2019nqd}
\begin{equation}
    \begin{gathered}
  {L^{\mu \nu }}{H_{\mu \nu }} =  
  {L_{\mu '\nu '}}{g^{\mu '\mu }}{g^{\nu '\nu }}{H_{\mu \nu }} \hfill \\
   = {L_{\mu '\nu '}}{\epsilon}^{\mu '}({\lambda _W}){\epsilon}_{}^{{\text{\dag }}\mu }({{\lambda ''}_W}){g_{{\lambda _W},{{\lambda ''}_W}}}{\epsilon}^{{\text{\dag }}\nu '}({{\lambda '}_W}){\epsilon}_{}^\nu ({{\lambda '''}_W}){g_{{{\lambda '}_W},{{\lambda '''}_W}}}{H_{\mu \nu }} \hfill \\
   = L({\lambda _W},{{\lambda '}_W}){g_{{\lambda _W},{{\lambda ''}_W}}}{g_{{{\lambda '}_W},{{\lambda '''}_W}}}H({{\lambda ''}_W},{{\lambda '''}_W}), \hfill \\ 
\end{gathered} 
\end{equation}
with the leptonic and hadronic tensors,
\begin{equation}
\label{Eq: leptonic tensor}
L({\lambda _W},{\lambda '_W}) = \epsilon^{\mu '}({\lambda _W})\epsilon^{{\text{\dag }}\nu '}({\lambda '_W}){L_{\mu '\nu '}},\    
\end{equation}
and 
\begin{equation}
H({\lambda _W},{\lambda '_W}) = {\epsilon}^{{\text{\dag }}\mu }({\lambda _W}){\epsilon}^\nu ({\lambda '_W}){H_{\mu \nu }}.
\end{equation}
In the initial $B_s$ rest frame, the momentum and polarization vectors are defined as
\begin{equation}
    \begin{gathered}
  p_1^\mu  = ({m_1},0), \hfill \\
  p_2^\mu  = ({E_2},0,0, - |{{\vec p}_2}|), \hfill \\
  {q^\mu } = ({q_0},0,0,|{{\vec p}_2}|), \hfill \\ 
\end{gathered} 
\end{equation}
with $E_2$ the energy of the $D_s$ meson, $q_0 = (m_1^2 - m_2^2 + q^2)/2m_1$ and \cite{Ivanov:2015tru}
\begin{equation}
\label{eq: pol pseu}
    \epsilon^\mu(t)=\frac{1}{{\sqrt {{q^2}} }}({q_0},0,0,|{{\vec p}_2}|),
    \epsilon^\mu(\pm) = \frac{1}{{\sqrt 2 }}(0, \mp 1, - i,0),
    \epsilon^\mu(0)=\frac{1}{{\sqrt {{q^2}} }}(|{{\vec p}_2}|,0,0,{q_0}).
\end{equation}
Applying Eqs. (\ref{eq: pseu form}) and (\ref{eq: pol pseu}), one obtains the helicity amplitudes with respect of form factors
\begin{equation}
    H_t = \frac{1}{\sqrt{q^2}}(m_1^2 - m_2^2)F_0(q^2),\
    H_{\pm} = 0,\
    H_0 = \frac{1}{\sqrt{q^2}}(2m_1|\vec{p_2}|)F_1(q^2).
\end{equation}
For the transition of $B_s \to D_s^*$, the hadronic tensor can be written as \cite{Ivanov:2019nqd}
\begin{equation}
\label{eq: hadronic tens vec}
     H({\lambda _W},{{\lambda '}_W}) = \epsilon^{\dag \mu} (\lambda_W) \epsilon^\nu({\lambda '}_W)H_{\mu\nu}=H_{\lambda_W \lambda_V}H^\dag_{\lambda '_W \lambda_V},
\end{equation}
where $H_{\lambda_W \lambda_V}$ is considered as 
\begin{equation}
    H_{\lambda _W,\lambda _V} = \epsilon^{\dag \mu} (\lambda _W)\epsilon^{\dag \alpha} _2(\lambda_V) T_{\mu \alpha }.
\end{equation}
The angular momentum conservation requires that ${\lambda _V} = {\lambda _W},{{\lambda '}_V} = {{\lambda '}_W}$ when ${\lambda _W},{{\lambda '}_W} =  \pm ,0$ and ${\lambda _V},{{\lambda '}_V} = 0$ while ${\lambda _W},{{\lambda '}_W} =  t$. In the initial meson rest frame, the polarization four-vectors of $D_s^*$ vector meson $\epsilon_2^\mu(\lambda_V)$ can be determined by
\begin{equation}
\label{eq:pol vec}
   \epsilon^\mu_2(\pm) = \frac{1}{{\sqrt 2 }}(0, \pm 1, - i,0),
    \epsilon^\mu_2(0)=\frac{1}{ {{m_2}} }(|{{\vec p}_2}|,0,0,{-E_2}).
\end{equation}
with the helicity components basis. Employing Eqs. (\ref{eq: vector amp}) and (\ref{eq:pol vec}), one finds the corresponding helicity amplitudes of $B_s \to D_s^*$
\begin{equation}
\label{Eq: helicity Bs Ds*}
    \begin{gathered}
  {H_t} =  - \frac{{2{m_1}|{{\vec p}_2}|}}{{\sqrt {{q^2}} }}{A_0}({q^2}), \hfill \\
  {H_ \pm } =  - ({m_1} + {m_2}){A_1}({q^2}) \pm \frac{{2{m_1}|{{\vec p}_2}|}}{{{m_1} + {m_2}}}V({q^2}) \hfill \\
  {H_0} =  - \frac{{{m_1} + {m_2}}}{{2{m_2}\sqrt {{q^2}} }}(m_1^2 - m_2^2 - {q^2}){A_1}({q^2}) + \frac{{2m_1^2|{{\vec p}_2}{|^2}}}{{({m_1} + {m_2})}(m_2\sqrt{q^2})}{A_2}({q^2}). \hfill \\ 
\end{gathered} 
\end{equation}
\subsection{Twofold angular distribution}
\label{section: decay width}
The twofold differential decay distribution versus the momentum transfer $q^2$ and polar angle $\theta$, which is the angle between the momentum of lepton and the momentum of daughter's meson in the rest frame of $W_\text{off-shell}$, can be written as
 
\begin{equation}
    \begin{gathered}
  \frac{{{d^2}\Gamma ({B _s} \to D _s^{(*)}{l^ + }{\nu _l})}}{{d{q^2}d\cos \theta }} 
   = \frac{{G_F^2|{V_{cb}}{|^2}}}{{{{(2\pi )}^3}}}\frac{{|{\vec p_2}|}}{{64m_1^2}}\left( {1 - \frac{{m_\ell^2}}{{{q^2}}}} \right){H^{\mu \nu }}{L_{\mu \nu }}, \hfill \\ 
\end{gathered} 
\end{equation}
where hadronic tensor parts can be obtained by Eqs. (\ref{eq: hadronic tens pseu}, \ref{eq: hadronic tens vec}) for $B_s \to D_s$ and $B_s \to D_s^{(*)}$ respectively. In the $W$ rest frame, leptonic tensor can be explained as
\begin{equation}
    {L_{\mu \nu }} = Tr[({\slashed{k}_1} + {m_l}){\mathcal{O}_\mu }{\slashed{k}_2}{\mathcal{O}_\nu }],
\end{equation}
for ${W^ - } \to {\ell ^ - }{\bar \nu _\ell }$, and
\begin{equation}
    {L_{\mu \nu }} = Tr[({\slashed{k}_1} - {m_l}){\mathcal{O}_\nu }{\slashed{k}_2}{\mathcal{O}_\mu }],
\end{equation}
for the case of ${W^ + } \to {\ell ^ + }{ \nu _\ell }$. They can be written in the following formula
\begin{equation}
\label{eq: Lmunu}
    {L_{\mu \nu }} = 8(k_1^\mu k_2^\nu  + k_1^\nu k_2^\mu  - {k_1}.{k_2}{g^{\mu \nu }} \pm i{\varepsilon ^{\mu \nu \alpha \beta }}{k_{1\alpha }}{k_{2\beta }}),
\end{equation}
where
\begin{equation}
    \begin{gathered}
  {q^\mu } = \left( {\sqrt {{q^2}} ,0,0,0} \right), \hfill \\
  k_1^\mu  = \left( {{E_1},|\vec{k_1}|\sin \theta \cos \chi ,|\vec{k_1}|\sin \theta \sin \chi ,|\vec{k_1}|\cos \theta } \right), \hfill \\
  k_2^\mu  = \left( {|\vec{k_1}|, - |\vec{k_1}|\sin \theta \cos \chi , - |\vec{k_1}|\sin \theta \sin \chi , - |\vec{k_1}|\cos \theta } \right), \hfill \\ 
\end{gathered} 
\end{equation}
with the energy ${E_1} = \frac{{{q^2} + m_\ell^2}}{{2\sqrt {{q^2}} }}, $ and three-momentum of the charged lepton $ |\vec{k_1}| = \frac{{{q^2} - m_\ell^2}}{{2\sqrt {{q^2}} }}$.
We need to define the polarization vectors in the $W$ rest frame to find the leptonic tensor elements. They have longitudinal, transverse, and time helicity components as
\begin{equation}
\label{eq: pol W rest}
    \begin{gathered}
  \epsilon^\mu ( \pm 1) = \frac{1}{{\sqrt 2 }}\left( {0, \mp 1, - i,0} \right), \hfill \\
  \epsilon^\mu (0) = \left( {0,0,0,1} \right), \hfill \\
  \epsilon^\mu (t) = \left( {1,0,0,0} \right). \hfill \\ 
\end{gathered} 
\end{equation}
Substituting Eqs.(\ref{Eq: leptonic tensor}, \ref{eq: pol W rest}) in Eq. (\ref{eq: Lmunu}), one gets
\begin{equation}
\label{Eq: leptonic matrix}
    \begin{gathered}
  {(2{q^2}v)^{ - 1}}L({\lambda _W},{{\lambda '}_W})(\theta ,\chi ) \hfill \\
   = \left( {\begin{array}{*{20}{c}}
  0&0&0&0 \\ 
  0&{{{(1 \mp \cos \theta )}^2}}&{ \mp \frac{2}{{\sqrt 2 }}(1 \mp \cos \theta )\sin \theta {e^{i\chi }}}&{{{\sin }^2}\theta {e^{2i\chi }}} \\ 
  0&{ \mp \frac{2}{{\sqrt 2 }}(1 \mp \cos \theta )\sin \theta {e^{ - i\chi }}}&{2{{\sin }^2}\theta }&{ \mp \frac{2}{{\sqrt 2 }}(1 \pm \cos \theta )\sin \theta {e^{i\chi }}} \\ 
  0&{{{\sin }^2}\theta {e^{ - 2i\chi }}}&{ \mp \frac{2}{{\sqrt 2 }}(1 \pm \cos \theta )\sin \theta {e^{ - i\chi }}}&{{{(1 \pm \cos \theta )}^2}} 
\end{array}} \right) \hfill \\
   + {\delta _\ell}\left( {\begin{array}{*{20}{c}}
  4&{ - \frac{4}{{\sqrt 2 }}\sin \theta {e^{ - i\chi }}}&{4\cos \theta }&{\frac{4}{{\sqrt 2 }}\sin \theta {e^{i\chi }}} \\ 
  { - \frac{4}{{\sqrt 2 }}\sin \theta {e^{i\chi }}}&{2{{\sin }^2}\theta }&{ - \frac{2}{{\sqrt 2 }}\sin 2\theta {e^{i\chi }}}&{ - 2{{\sin }^2}\theta {e^{2i\chi }}} \\ 
  {4\cos \theta }&{ - \frac{2}{{\sqrt 2 }}\sin 2\theta {e^{ - i\chi }}}&{4{{\cos }^2}\theta }&{\frac{2}{{\sqrt 2 }}\sin 2\theta {e^{i\chi }}} \\ 
  {\frac{4}{{\sqrt 2 }}\sin \theta {e^{ - i\chi }}}&{ - 2{{\sin }^2}\theta {e^{ - 2i\chi }}}&{\frac{2}{{\sqrt 2 }}\sin 2\theta {e^{ - i\chi }}}&{2{{\sin }^2}\theta } 
\end{array}} \right), \hfill \\ 
\end{gathered} 
\end{equation}
with velocity-type $v = 1 - m_\ell^2/{q^2}$ and ${\delta _\ell} = m_\ell^2/2{q^2}$ helicity-flip factors, and the upper/lower sign refers to the two configurations ${\ell^ - }{\bar \nu _\ell}/{\ell^ + }{\nu _\ell}$. Integrating over the azimuthal angle $\chi$, Eq. (\ref{Eq: leptonic matrix}) becomes
\begin{equation}
    \begin{gathered}
  \frac{1}{2\pi}{(2{q^2}v)^{ - 1}}L({\lambda _W},{{\lambda '}_W})(\theta ) \hfill \\
   = \left( {\begin{array}{*{20}{c}}
  0&0&0&0 \\ 
  0&{{{(1 \mp \cos \theta )}^2}}&0&0 \\ 
  0&0&{2{{\sin }^2}\theta }&0 \\ 
  0&0&0&{{{(1 \pm \cos \theta )}^2}} 
\end{array}} \right) \hfill \\
   + {\delta _\ell }\left( {\begin{array}{*{20}{c}}
  4&0&{4\cos \theta }&0 \\ 
  0&{2{{\sin }^2}\theta }&0&0 \\ 
  {4\cos \theta }&0&{4{{\cos }^2}\theta }&0 \\ 
  0&0&0&{2{{\sin }^2}\theta } 
\end{array}} \right). \hfill \\ 
\end{gathered} 
\end{equation}
We continue with the lower sign in the matrix since it is our study case $B_s \to D_s^{(*)}{\ell ^ + }{\nu _\ell }$. Therefore, the twofold polar angular distribution becomes \cite{Ivanov:2015tru}
\begin{equation}
    \begin{gathered}
  \frac{{{d}\Gamma ({B _s} \to {D _s^{(*)}}{l^ + }{\nu _\ell})}}{{d{q^2}d\cos \theta }} = \frac{{G_F^2|{V_{cb}}{|^2}|{\vec{p_2}}|{q^2}{v^2}}}{{32{{(2\pi )}^3}m_1^2}} \times \{ (1 + {\cos ^2}\theta ){\mathcal{ H}_U} + 2{\sin ^2}\theta {\mathcal{ H}_L} + 2\cos \theta {\mathcal{ H}_P} \hfill \\
   + 2{\delta _\ell}[{\sin ^2}\theta {\mathcal{ H}_U} + 2{\cos ^2}\theta \mathcal{ H}_L + 2\mathcal{ H}_S - 4\cos \theta \mathcal{ H}_{SL}]\},  \hfill \\ 
\end{gathered} 
\end{equation}
 with the following helicity structure functions
 \begin{equation}
     \begin{gathered}
  {\mathcal{H}_U} = |{H_ + }{|^2} + |{H_ - }{|^2},{\mathcal{H}_L} = |{H_0}{|^2},{\mathcal{H}_S} = |{H_t}{|^2}, \hfill \\
  {\mathcal{H}_P} = |{H_ + }{|^2} - |{H_ - }{|^2},{\mathcal{H}_{SL}} = \operatorname{Re} (H_0{H_t^\dag}). \hfill \\ 
\end{gathered} 
 \end{equation}
 Finally, the differential decay distribution over transfer momentum $q^2$, is defined as
\begin{equation}
\label{eq: decay unpol}
    \frac{{d\Gamma ({B _s} \to {D _s^{(*)}}{l^ + }{\nu _l})}}{{d{q^2}}} = \frac{{G_F^2|{V_{cb}}{|^2}|{\vec{p_2}}|{q^2}{v^2}}}{{12{{(2\pi )}^3}m_1^2}} \times {\mathcal{H}_{tot}},
\end{equation}
where 
\begin{equation}
    {\mathcal{H}_{tot}} = {\mathcal{H}_U} + {\mathcal{H}_L} + {\delta _\ell}[{\mathcal{H}_U} + {\mathcal{H}_L} + 3{\mathcal{H}_S}].
\end{equation}
\subsection{Physical observables}
\label{sec: observables}
In addition to the branching fraction of semileptonic decay $B_s \to D_s^{*}$, we can investigate other physical observables. The forward-backward asymmetry is given by \cite{Ivanov:2019nqd}
\begin{equation}
    A_{FB}^\ell ({q^2}) = \frac{{\int\limits_0^1 {d\cos \theta \frac{{d\Gamma }}{{d{q^2}d\cos \theta }}}  - \int\limits_{ - 1}^0 {d\cos \theta \frac{{d\Gamma }}{{d{q^2}d\cos \theta }}} }}{{\int\limits_0^1 {d\cos \theta \frac{{d\Gamma }}{{d{q^2}d\cos \theta }}}  + \int\limits_{ - 1}^0 {d\cos \theta \frac{{d\Gamma }}{{d{q^2}d\cos \theta }}} }} = \frac{3}{4}\frac{{{\mathcal{H_P}} - 4{\delta _\ell }{\mathcal{H}_{SL}}}}{{{\mathcal{H}_{tot}}}}.
\end{equation}
For the longitudinal polarization of the lepton with the following vector
\begin{equation}
    s_L^\mu  = \frac{1}{{{m_\ell }}}(|{{\vec k}_1}|,{E_1}\sin \theta ,0,{E_1}\cos \theta ),
\end{equation}
the leptonic tensor is defined as
\begin{equation}
    {L_{\mu \nu }}({s_L}) =  \mp ({s_{L\mu }}{k_{2\nu }} + {s_{L\nu }}{k_{2\mu }} - {s_L}.{k_2}{g_{\mu \nu }} \pm i{\varepsilon _{\mu \nu \alpha \beta }}s_L^\alpha k_2^\beta ),
\end{equation}
where the upper/lower sign corresponds to the ${\ell ^ - }{{\bar \nu }_\ell }/{\ell ^ + }{\nu _\ell }$.
Then the polarized differential semileptonic decay can be written as
\begin{equation}
    \frac{{d\Gamma (s_L)}}{{d{q^2}}} = \frac{{G_F^2|{V_{cb}}{|^2}|{\vec{p_2}}|{q^2}{v^2}}}{{12{{(2\pi )}^3}m_1^2}} \times [{\mathcal{H}_{U}+\mathcal{H}_{L}-\delta _\ell (\mathcal{H}_{U}+\mathcal{H}_{L}+3\mathcal{H}_{S})}].
\end{equation}
The ratio of polarized decay distribution to the unpolarized one, Eq. (\ref{eq: decay unpol}), is known as longitudinal polarization of the lepton which can be written \cite{Ivanov:2019nqd}
\begin{equation}
   P_L^\ell(q^2) = \frac{{\mathcal{H}_{U}+\mathcal{H}_{L}-\delta _\ell (\mathcal{H}_{U}+\mathcal{H}_{L}+3\mathcal{H}_{S})}}{\mathcal{H}_{tot}},
\end{equation}
for the case ${W^ + } \to {\ell ^ + }{\nu _\ell }$. To get the transverse polarization $P_T^\ell$, one can consider $s_T^\mu  = (0,\cos \theta ,0, - \sin \theta )$, and the transverse polarization of lepton becomes \cite{Ivanov:2019nqd} 
\begin{equation}
   P_T^\ell(q^2) =\frac{-3\pi\sqrt{\delta _\ell}}{4\sqrt{2}} \frac{{(\mathcal{H}_{P}+2\mathcal{H}_{SL})}}{\mathcal{H}_{tot}}.
\end{equation}
Furthermore, the lepton-side convexity parameter can be defined as \cite{Ivanov:2019nqd}
\begin{equation}
    C_F^\ell ({q^2}) = \frac{3}{4}(1 - 2{\delta _\ell })\frac{{{\mathcal{H}_U} - 2{\mathcal{H}_L}}}{{{\mathcal{H}_{tot}}}}.
\end{equation}
The longitudinal polarization fractions of the $D_s^{(*)}$ meson is given by \cite{Ivanov:2019nqd}
\begin{equation}
   F_L^\ell ({q^2}) = \frac{{(1 + {\delta _\ell)\mathcal{H}_L+3\delta _\ell\mathcal{H}_S} }}{{{\mathcal{H}_{tot}}}}.
\end{equation}
\subsection{$B_s \to D_sP$ and $B_s \to D_sV$ }
\label{sec: twobody}
The factorization assumption based on the vacuum saturation approximation works reasonably well in analyzing
heavy-quark physics. In the factorization approach, the effective Hamiltonian of ${B_s} \to D_sP(V)$ can be written as
\begin{equation}
    {\mathcal{H}_{{\text{eff}}}} = \frac{{{G_F}}}{{\sqrt 2 }}{V_{cb}}{V_{{q_1}{q_2}}}({C_1}{O_1} + {C_2}{O_2}) + {\text{H}}{\text{.c}}{\text{.}},
\end{equation}
where ${V_{{q_1}{q_2}}}$ is the corresponding CKM element, $C_{(1,2)}$ are the Wilson coeﬃcients and the local tree four-quark operators are given by \cite{Sun:2023uyn}
\begin{equation}
    \begin{gathered}
  {O_1} = {\bar q}_{1,\alpha }{\gamma _\mu }(1 - {\gamma _5}){c_\alpha}.{{\bar u}_{\beta }}{\gamma ^\mu }(1 - {\gamma _5})q_{2,\beta }, \hfill \\
   {O_2} = {\bar q}_{1,\alpha }{\gamma _\mu }(1 - {\gamma _5}){c_\beta}.{{\bar u}_{\beta }}{\gamma ^\mu }(1 - {\gamma _5})q_{2,\alpha }, \hfill \\ 
\end{gathered} 
\end{equation}
with $\alpha, \beta$ color indices. Using the effective Hamiltonian, the matrix elements for the decays ${B_s} \to D_sP(V)$ become as
\begin{equation}
    \begin{gathered}
  \mathcal{A}({B_s} \to D_s^{}P(V)) = \left\langle {D_s^{}P(V)|\mathcal{H}_{\text{eff}}}|{B_s} \right\rangle  \hfill \\
   = \frac{{{G_F}}}{{\sqrt 2 }}{V_{cb}}{V_{{q_1}{q_2}}}{a_1}\left\langle {P(V)|{J^\mu }|0} \right\rangle \left\langle {{D_s}|{J_\mu }|{B_s}} \right\rangle,  \hfill \\ 
\end{gathered} 
\end{equation}
where $a_1$ is a QCD factor, defined in terms of the Wilson coefﬁcients as $a_1 = C_2 + C_1/3$. One can write $\left\langle {P|{J^\mu }|0} \right\rangle  =  - i{f_P}{q_\mu }$ and $\left\langle {V|{J^\mu }|0} \right\rangle  = {f_V}{m_V}\epsilon _\mu ^*$ in terms of decay constants of pseudoscalar and vector mesons. Thus, the amplitude for ${B_s} \to D_sP$ and ${B_s} \to D_sV$ can be specified as follows:
\begin{equation}
 \begin{gathered}
    \mathcal{A}({B_s} \to D_s^{}P) = \frac{{{G_F}}}{{\sqrt 2 }}{V_{cb}}{V_{{q_1}{q_2}}}{a_1}(m_1^2 - m_2^2){f_P}f_0^{{B_s}{D_s}}(m_P^2),\hfill \\
    \mathcal{A}({B_s} \to D_s^{}V) = \sqrt 2 {G_F}{V_{cb}}{V_{{q_1}{q_2}}}{a_1}{m_V}(\epsilon _V^*.{p_{{B_s}}}){f_V}f_ + ^{{B_s}{D_s}}(m_V^2),
    \end{gathered}
\end{equation}
with regard to the decay constants and transition form factors. The corresponding expressions for the decay width of these two-body decays can be written as \cite{Fu-Sheng:2011fji}
\begin{equation}
\begin{gathered}
    \Gamma ({B_s} \to D_s P) = \frac{p}{{8\pi m_1^2}}|\mathcal{A}{|^2},\hfill \\
    \Gamma ({B_s} \to D_s V) = \frac{p}{{8\pi m_1^2}}\sum\limits_{\text{pol}} {|\mathcal{A}{|^2}} 
    \end{gathered}
\end{equation}
with $p$ the momentum of either meson in the ﬁnal state,  $p = p_{D_s} = p_{P/V} = {\sqrt {(m_1^2 - {{({m_2} + {m_{P/V}})}^2})(m_1^2 - {{({m_2} - {m_{P/V}})}^2})} }/{{2m_1^{}}}$ in the center-of-mass frame. 

\subsection{Form factors }
\label{sec: form factors}
To calculate the physical observables of semileptonic decays, the transition form factors and their dependence on transfer momentum should be known. The transition hadronic form factors are also important in nonleptonic decays. There are numerous models available to describe them. Some of them such as perturbative predictions are limited to a region of small values of $q^2$. Lattice QCD results are also reliable close to zero recoil, the reason originating from avoiding large statistical errors. In the current work, we make the extrapolation by using two formulations, dipole and exponential functions as follows \cite{Fu-Sheng:2011fji, Wang:2014yia, Sun:2023uyn}
\begin{equation}
    \begin{gathered}
  F({q^2}) = \frac{{F(0)}}{{1 - a_1\frac{{{q^2}}}{{m_{pole}^2}} + a_2\frac{{{q^4}}}{{m_{pole}^4}}}}, \hfill \\
  F({q^2}) = F(0)\exp [a_1{q^2} + a_2{q^4}], \hfill \\ 
\end{gathered} 
\end{equation}
where $F(0), a_1, a_2$ are free parameters in both equations obtained by performing fit. $F(q^2)$ stands for the form factors ${f_{0, + }}(q^2),{A_{0,1,2}}(q^2),V(q^2)$ and $m_{pole}$ is the mass of the pole. The input parameters are identified in the next section.
\section{Numerical Analysis}
\label{section: numeric}
\begin{figure}
     \centering
     \begin{subfigure}[b]{0.49\textwidth}
         \centering
         \includegraphics[width=\textwidth]{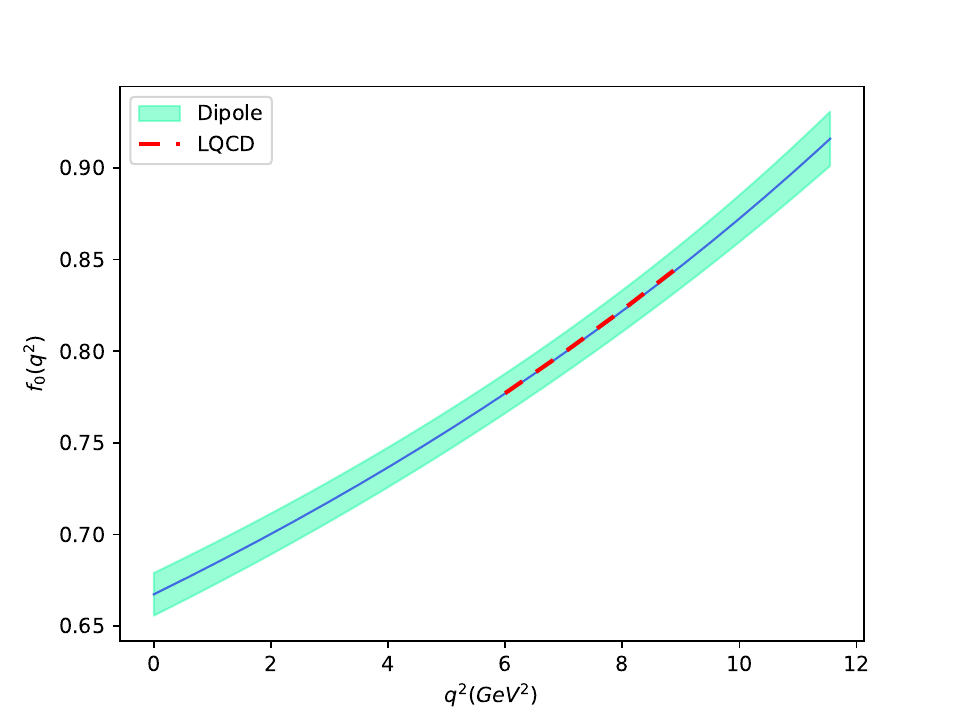}
     \end{subfigure}
     \hfill
     \begin{subfigure}[b]{0.49\textwidth}
         \centering
         \includegraphics[width=\textwidth]{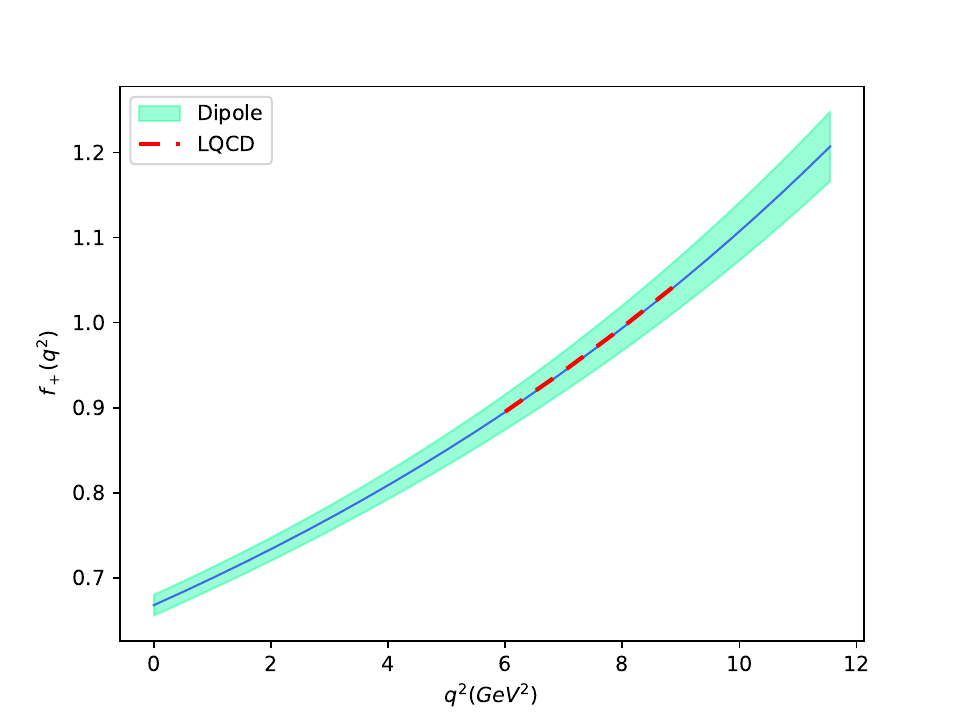}
        
     \end{subfigure}
     \caption{The fitted form factors $f_0$ and $f_+$ versus $q^2$ using dipole model.}
     \hfill
     \label{BsDs_dipole}
\end{figure}

\begin{figure}
     \centering
     \begin{subfigure}[b]{0.49\textwidth}
         \centering
         \includegraphics[width=\textwidth]{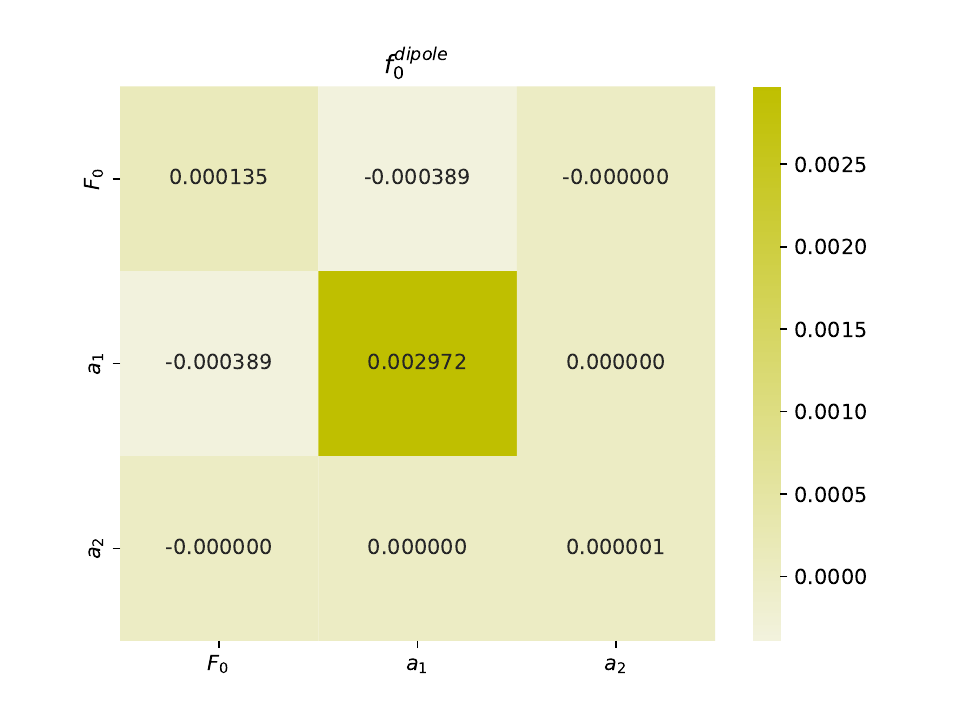}
     \end{subfigure}
     \hfill
     \begin{subfigure}[b]{0.49\textwidth}
         \centering
         \includegraphics[width=\textwidth]{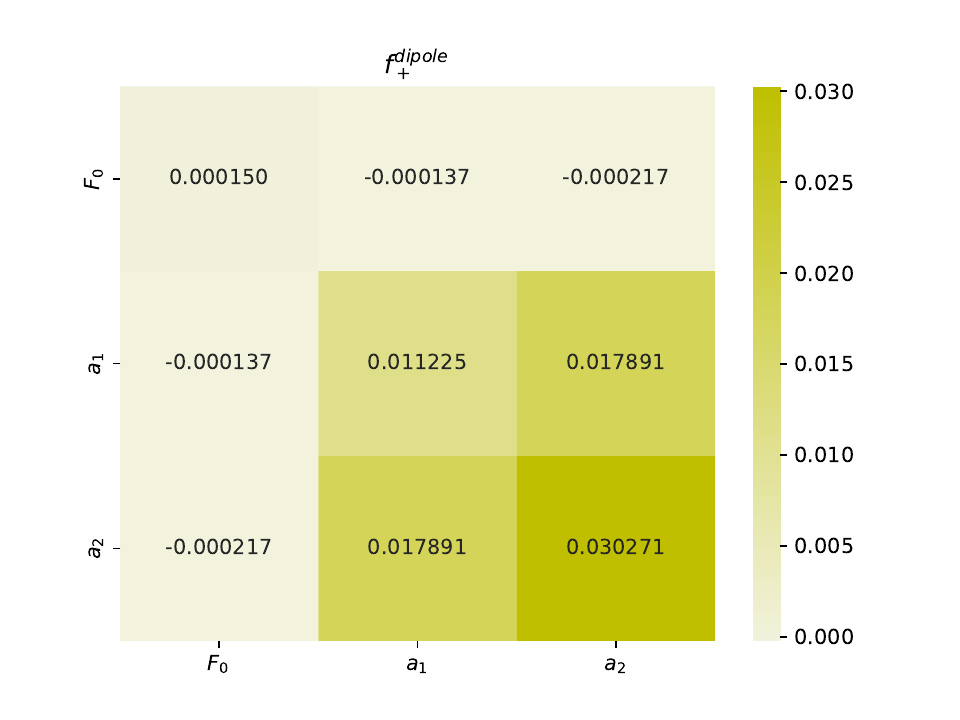}
        
     \end{subfigure}
     \caption{The covariance of parameters in the dipole form factors $f_0$ and $f_+$.}
     \hfill
     \label{BsDs_cov_dipole}
\end{figure}

\begin{figure}
     \centering
     \begin{subfigure}[b]{0.49\textwidth}
         \centering
         \includegraphics[width=\textwidth]{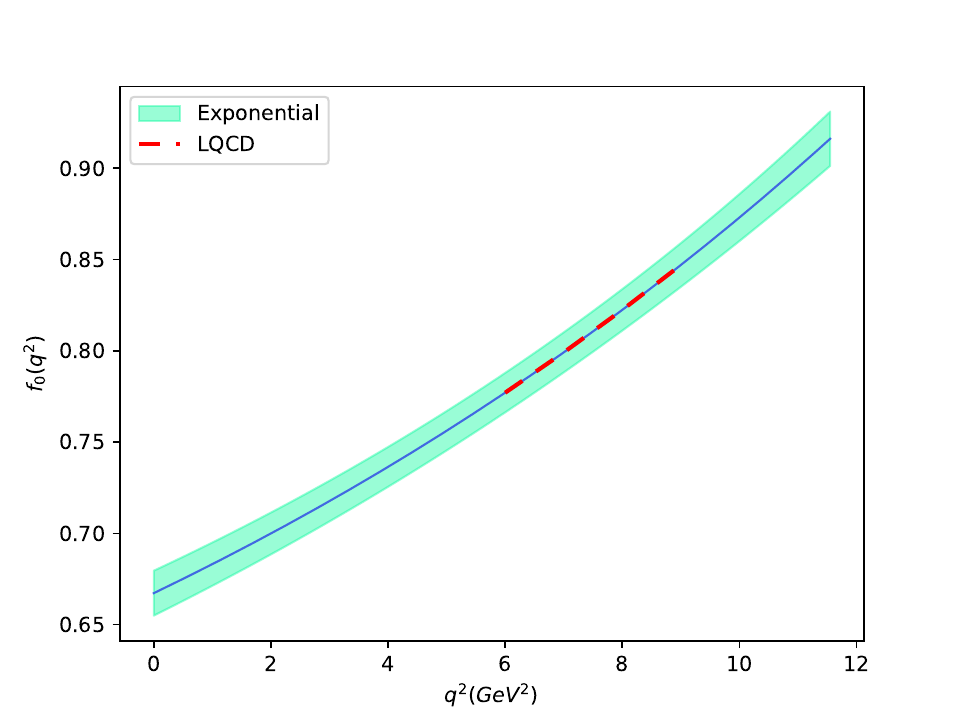}
     \end{subfigure}
     \hfill
     \begin{subfigure}[b]{0.49\textwidth}
         \centering
         \includegraphics[width=\textwidth]{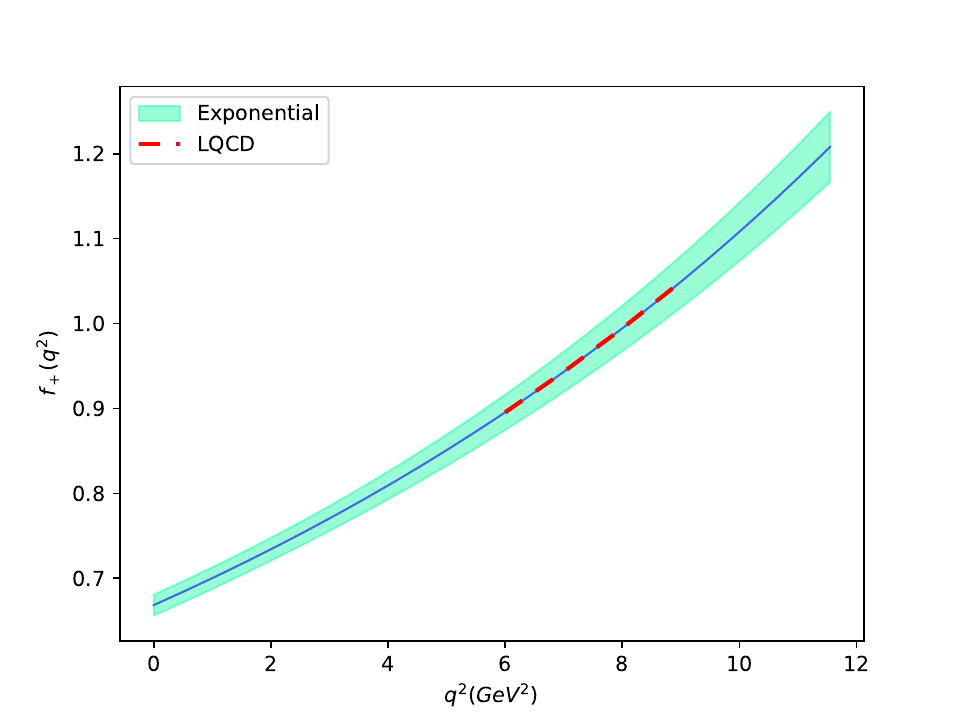}
        
     \end{subfigure}
     \caption{The fitted form factors $f_0$ and $f_+$ versus $q^2$ using exponential model.}
     \hfill
     \label{BsDs_exp}
\end{figure}

\begin{figure}
     \centering
     \begin{subfigure}[b]{0.49\textwidth}
         \centering
         \includegraphics[width=\textwidth]{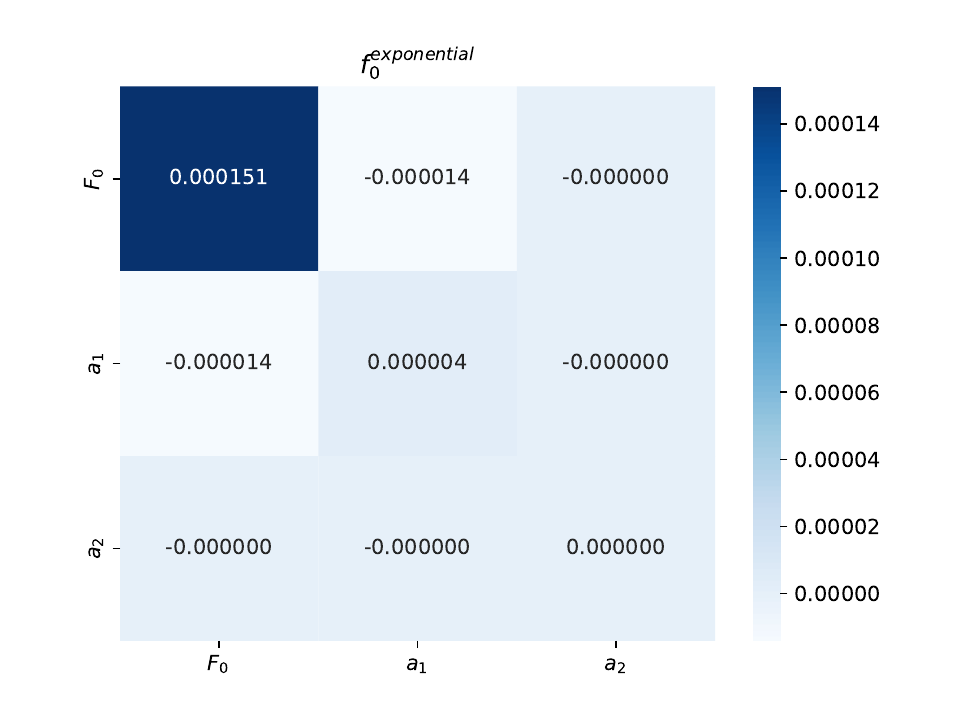}
     \end{subfigure}
     \hfill
     \begin{subfigure}[b]{0.49\textwidth}
         \centering
         \includegraphics[width=\textwidth]{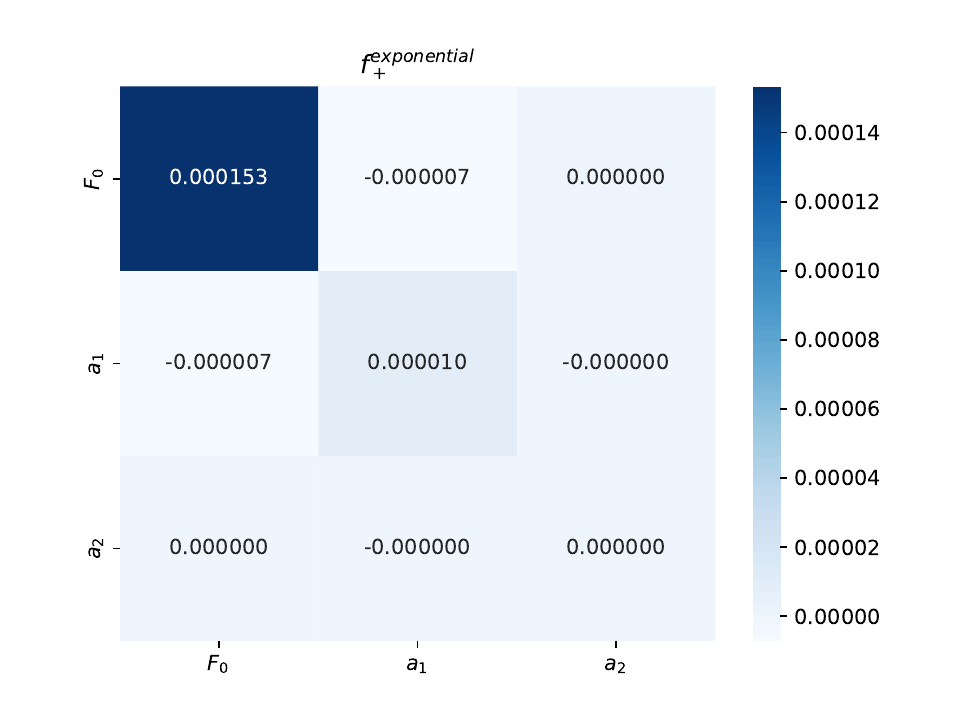}
        
     \end{subfigure}
     \caption{The covariance of parameters in the exponential form factors $f_0$ and $f_+$.}
     \hfill
     \label{BsDs_cov_exp}
\end{figure}

In this section, we first discuss $B_s \to D_s \ell \nu$ and its numerical observables. After that, $B_s \to D_s^* \ell \nu$ are investigated. In the computational evaluations, the following inputs are employed \cite{ParticleDataGroup:2024cfk}:
\begin{center}
    $\begin{gathered}
  {m_{{B_{c0}}}} = 6.704\,\,{\text{GeV, }}{m_{B_c^*}} = 6.329\,\,{\text{GeV, }}{m_{{B_s}}} = 5.3669\,\,{\text{GeV,}} \hfill \\
  {m_{{D_s}}} = 1.9683\,\,{\text{GeV}},\,\,{m_e} = 0.511\,\,{\text{MeV,}}\,\,{m_\mu } = 105.658\,\,{\text{MeV,}} \hfill \\
  {m_\tau } = 1.777\,\,{\text{GeV}},\,\,{\tau _{{B_s}}} = 1.527\,\,{\text{ps,}}\,\,\,{m_{D_s^*}} = 2.112\,\,{\text{GeV}}, \hfill \\
  {f_{{D_s}}} = 249.9\,\,{\text{MeV}},\,\,{f_{{K^ + }}} = 155.7\,\,{\text{MeV}},\,\,{f_{{K^*}}} = 0.217\,\,{\text{GeV}}, \hfill \\
  {f_\rho } = 0.209\,\,{\text{GeV}},\,\,{f_{D_s^*}} = {\text{0}}{\text{.315  GeV,}}\,\,{f_{{\pi ^ + }}} = {\text{130}}{\text{.2 MeV,}} \hfill \\
  {V_{cb}} = 0.0408,\,\,{V_{cs}} = 0.975,\,\,{V_{ud}} = 0.974,\,\,{V_{us}} = 0.224. \hfill \\ 
\end{gathered} $
\end{center}
In Ref. \cite{McLean:2019qcx}, HPQCD collaboration presented a lattice QCD determination of the $B_s \to D_s \ell \nu$ using the BCL parametrization \cite{Bourrely:2008za} for the form factors. We used their mean values and the covariance matrix associated with the parameters in the BCL formulation to perform the fitting in the transition of $B_s \to D_s \ell \nu$. Our results are shown in Figs. \ref{BsDs_dipole}, \ref{BsDs_cov_dipole}, \ref{BsDs_exp}, \ref{BsDs_cov_exp}  for the dipole and exponential models respectively. Figs. \ref{BsDs_cov_dipole} and \ref{BsDs_cov_exp} show the covariance  between our obtained parameters for $F(0), a_1$ and $a_2$. In the dipole form, they are larger. In Table \ref{tab: fit_params_BsDs}, we present the calculated parameters $F(0), a_1$, and $a_2$ for $f_0$ and $f_+$ in both models. As one can see, the values of $F(0)$ are near in both formulations, however, $a_1$ and $a_2$ have more different values. The average of ${\chi ^2}$ in the dipole model is obtained as $0.1437$ and equal to $0.0483$ for the exponential one. By using the obtained parameters in Table \ref{tab: fit_params_BsDs}, we tabulate the decay rates of semileptonic transitions of $B_s \to D_s \ell \nu$ in Table \ref{tab: BsDs_BR}. Our predictions of $BR({B_s} \to {D_s}\mu \nu )$, $BR({B_s} \to {D_s}e \nu )$, and $BR({B_s} \to {D_s}\tau \nu )$ are close to the QCD sum rules approaches, $BR({B_s} \to {D_s}\ell \nu ) = 2.03_{ - 0.49}^{ + 0.35}$ \cite{Zhang:2021wnv} and $BR({B_s} \to {D_s}\tau \nu ) = 0.606_{ - 0.211}^{ + 0.266}$ \cite{Zhang:2022opp}. The semi-muonic channel was reported by PDG as $BR(B_s^0 \to D_s^ - {\mu ^ + }{\nu _\mu }) = (2.31 \pm 0.21)\% $ \cite{ParticleDataGroup:2024cfk}. Our values, $(2.3065 \pm 0.9503)\% $ and $(2.3084 \pm 0.9513)\% $ are both in good agreement with them. By employing of equations related to physical observables in section \ref{sec: observables}, we can obtain forward-backward asymmetry, longitudinal, transverse lepton polarization, and leptonic convexity parameters for different leptonic channels of $B_s \to D_s \ell \nu$, see Tables \ref{tab: FB_BsDs} and \ref{tab: polar_BsDs}. The longitudinal polarization of the lepton $\tau$ had been reported as $0.30$ in PQCD approach \cite{Hu:2019bdf}. Our magnitude of this observable is near with them.

\begin{table}[htb]
\caption{Fitted parameters $F(0), a_1$, and $a_2$ for $f_0$ and $f_+$ in transition of $B_s \to D_s$.}
\label{tab: fit_params_BsDs}
\begin{tabularx}{1\textwidth}{>{\centering\arraybackslash}X | >{\centering\arraybackslash}X | >{\centering\arraybackslash}X |>{\centering\arraybackslash}X | >{\centering\arraybackslash}X}

\hline\hline

 & $f_0^{dipole}({q^2})$ & $f_0^{exponential}({q^2})$ & $f_ + ^{dipole}({q^2})$ & $f_ + ^{exponential}({q^2})$
\\
\hline
$F(0)$ & $0.6673 \pm 0.0116$ & $0.6672 \pm 0.0123 $ & $0.6682 \pm 0.0122$  & $0.6684 \pm 0.0124$
\\
\hline
$a_1$ & $1.0574 \pm 0.0545$ & $0.0232 \pm 0.0019$ & $1.8461 \pm 0.1059$ & $0.0460 \pm 0.0032$
\\
\hline
$a_2$ & $0.0040 \pm 0.0020$ & $0.0004 \pm 0.0000$ & $1.0333 \pm 0.1740$  & $0.0005 \pm 0.0000$
\\
\hline\hline
\end{tabularx}
\end{table}

\begin{table}[htb]
\caption{Decay widths in $10^{(-15)}$ GeV and branching ratios in percentage for semileptonic transition of $B_s \to D_s \ell \nu_{\ell}$.}
\label{tab: BsDs_BR}
\begin{tabularx}{1\textwidth}{>{\centering\arraybackslash}X | >{\centering\arraybackslash}X | >{\centering\arraybackslash}X |>{\centering\arraybackslash}X | >{\centering\arraybackslash}X}

\hline\hline

Mode & $\Gamma$ with dipole & $\Gamma$ with exponential & $BR$ with dipole  & $BR$ with exponential
\\
\hline
$B_s^0 \to D_s^ - {\mu ^ + }{\nu _\mu }$ & $9.9374 \pm 4.0941$ & $9.9455 \pm 4.0988$ & $2.3065 \pm 0.9503$  & $2.3084 \pm 0.9513$
\\
\hline
$B_s^0 \to D_s^ - {e^ + }{\nu _e}$ & $9.9704 \pm 0.0002$ & $9.9786 \pm 0.0002$ & $2.3142 \pm 0.0000$ & $2.3161 \pm 0.0000$
\\
\hline
$B_s^0 \to D_s^ - {\tau ^ + }{\nu _\tau }$ & $2.9643 \pm 0.0180$ & $2.9675 \pm 0.0180$ & $0.6880 \pm 0.0042$  & $0.6888 \pm 0.0042$
\\
\hline\hline
\end{tabularx}
\end{table}

\begin{table}[htb]
\caption{Forward-backward asymmetry for semileptonic transition of $B_s \to D_s \ell \nu_{\ell}$.}
\label{tab: FB_BsDs}
\begin{tabularx}{1\textwidth}{>{\centering\arraybackslash}X | >{\centering\arraybackslash}X | >{\centering\arraybackslash}X |>{\centering\arraybackslash}X }

\hline\hline

 & $\left\langle {\mathcal{ A} _{FB}^e} \right\rangle $ & $\left\langle {\mathcal{ A} _{FB}^\mu} \right\rangle $ & $\left\langle {\mathcal{ A} _{FB}^\tau} \right\rangle $
\\
\hline
dipole & $-0.5175 \times {10^{ - 6}}$ & $-0.0138$ & $-0.3604$  
\\
\hline
exponential & $-0.5171 \times {10^{ - 6}}$ & $-0.0138$ & $-0.3604$ 
\\
\hline\hline
\end{tabularx}
\end{table}

\begin{figure}
     \centering
     \begin{subfigure}[b]{0.49\textwidth}
         \centering
         \includegraphics[width=\textwidth]{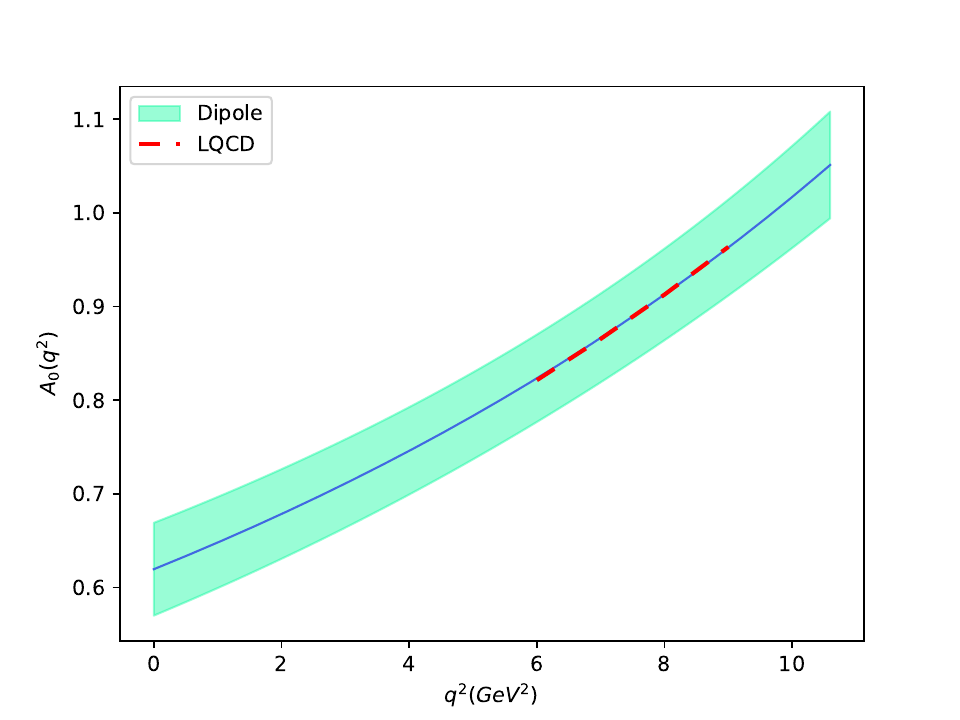}
     \end{subfigure}
     \hfill
     \begin{subfigure}[b]{0.49\textwidth}
         \centering
         \includegraphics[width=\textwidth]{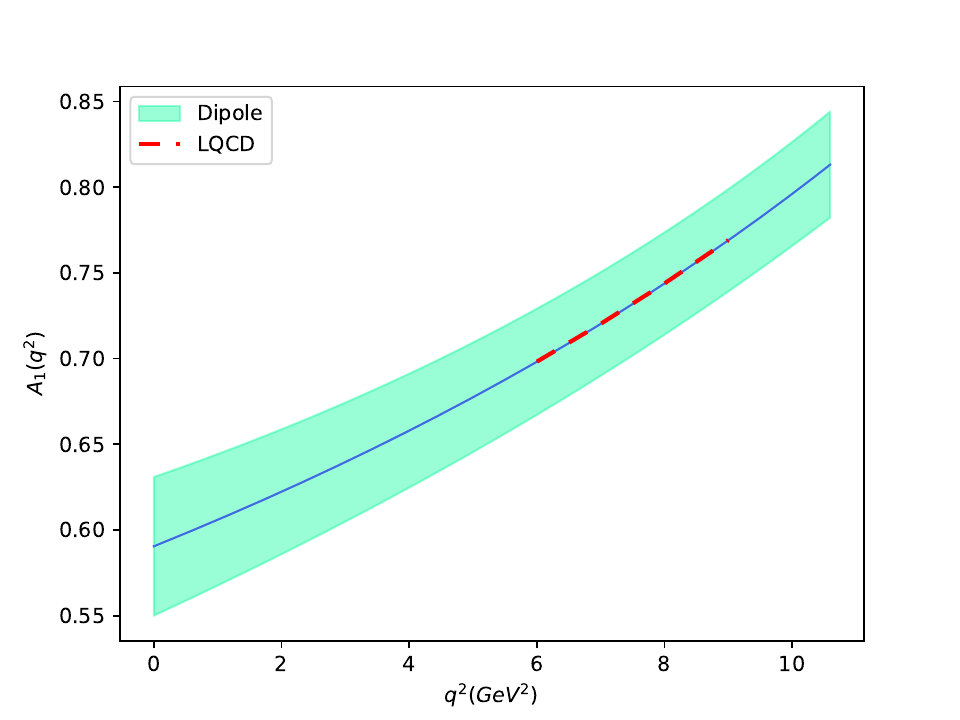}
     \end{subfigure}
     \hfill
     \begin{subfigure}[b]{0.49\textwidth}
         \centering
         \includegraphics[width=\textwidth]{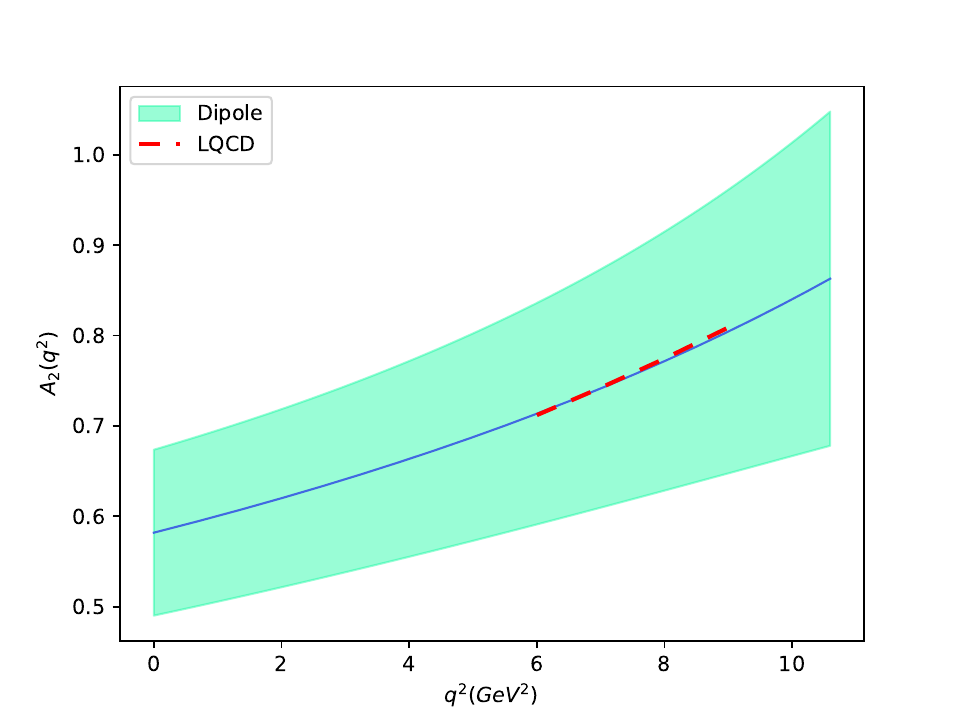}
     \end{subfigure}
     \hfill
          \begin{subfigure}[b]{0.49\textwidth}
         \centering
         \includegraphics[width=\textwidth]{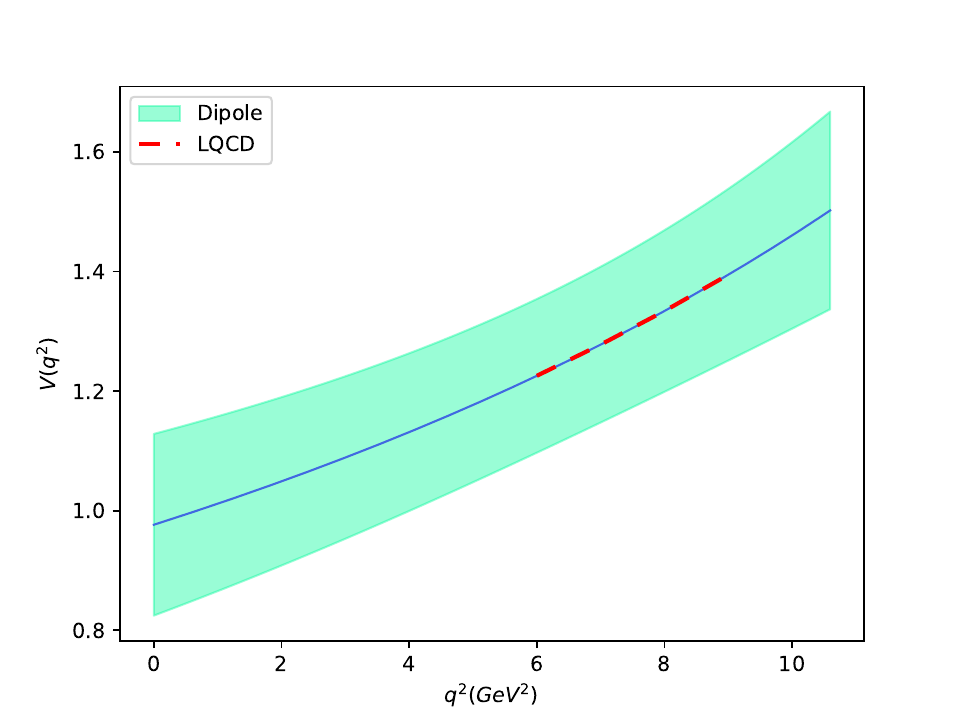}
     \end{subfigure}
     \hfill
             
      \caption{Form factor fits to LQCD data for transition $B_s \to D_s^*$ using dipole formula.} 
      \label{fig: fits BsDS* dipole}
\end{figure}

\begin{figure}
     \centering
     \begin{subfigure}[b]{0.49\textwidth}
         \centering
         \includegraphics[width=\textwidth]{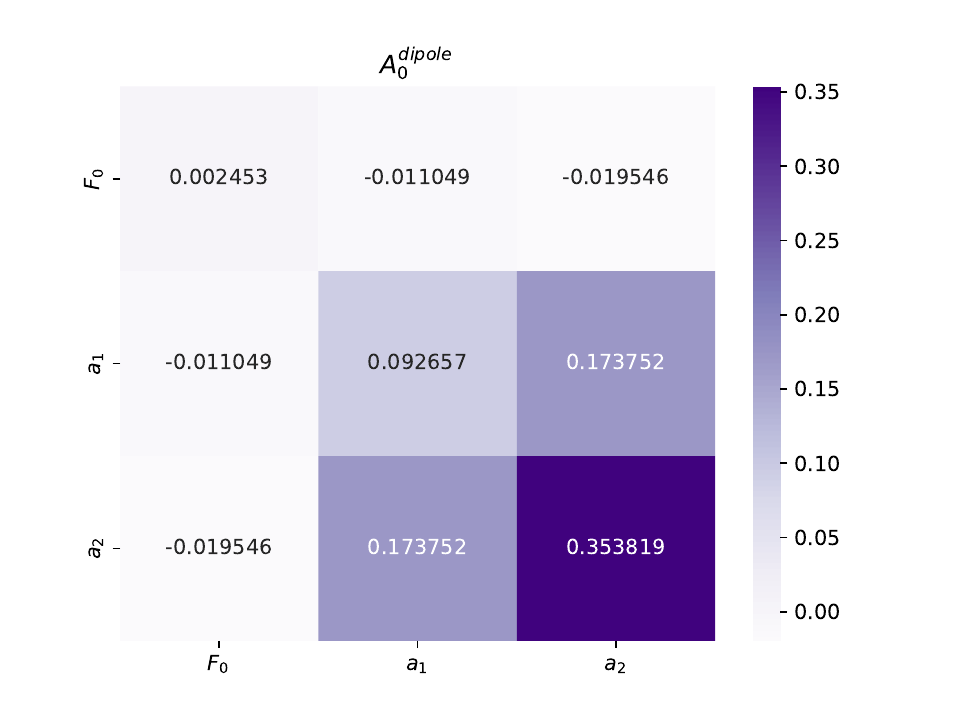}
     \end{subfigure}
     \hfill
     \begin{subfigure}[b]{0.49\textwidth}
         \centering
         \includegraphics[width=\textwidth]{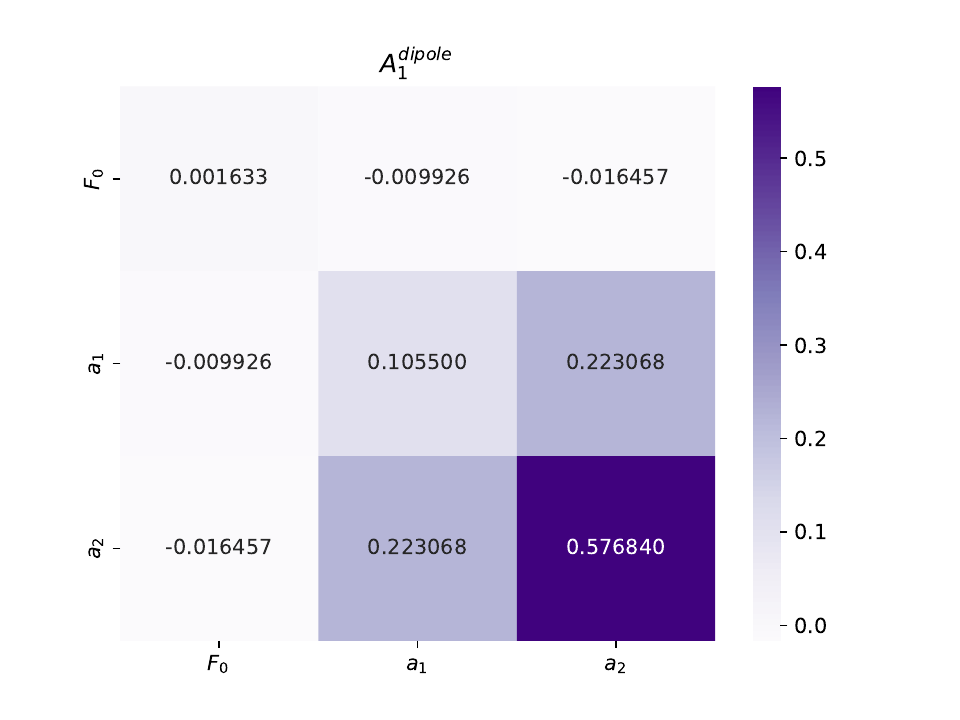}
     \end{subfigure}
     \hfill
     \begin{subfigure}[b]{0.49\textwidth}
         \centering
         \includegraphics[width=\textwidth]{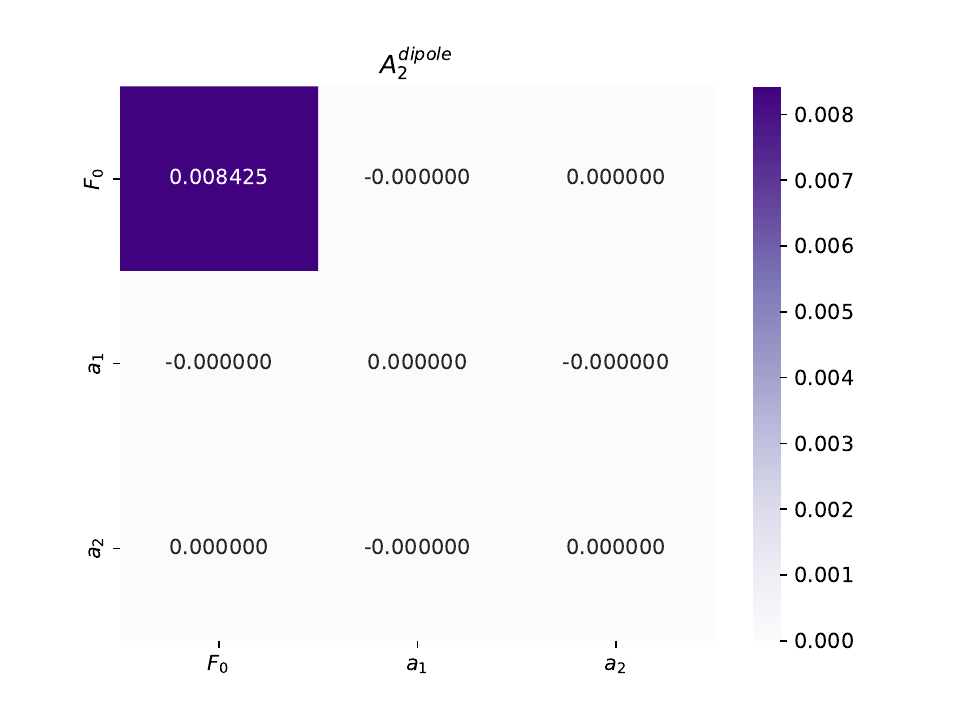}
     \end{subfigure}
     \hfill
          \begin{subfigure}[b]{0.49\textwidth}
         \centering
         \includegraphics[width=\textwidth]{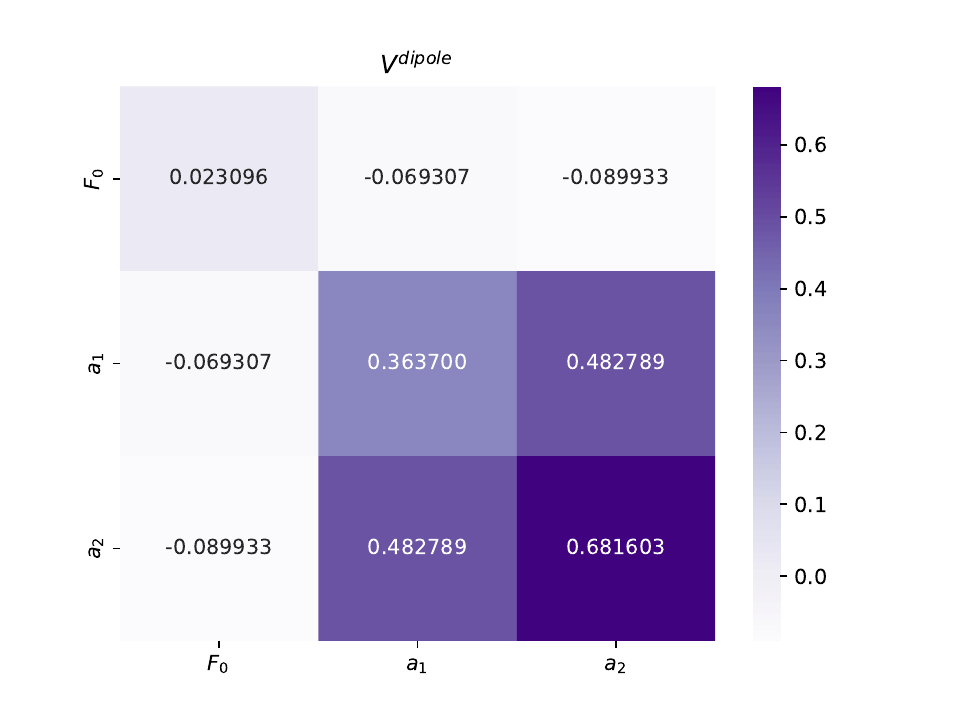}
     \end{subfigure}
     \hfill
             
      \caption{The covariance of parameters in the dipole form factors $V, A_0, A_1, A_2$.} 
      \label{fig: cov-dipole-BsDs*}
\end{figure}

\begin{figure}
     \centering
     \begin{subfigure}[b]{0.49\textwidth}
         \centering
         \includegraphics[width=\textwidth]{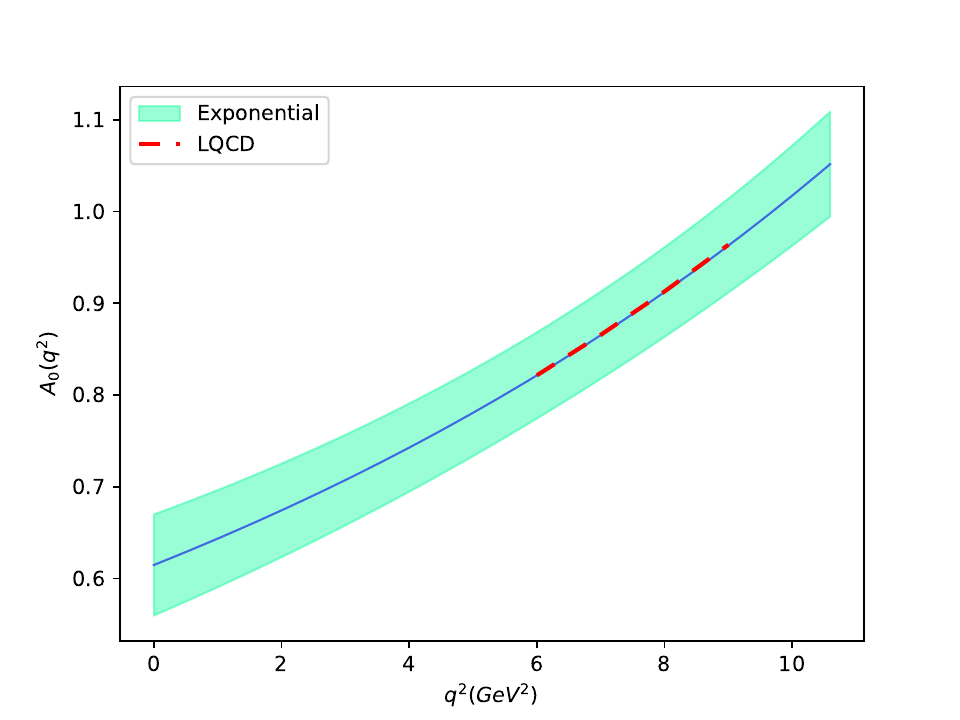}
     \end{subfigure}
     \hfill
     \begin{subfigure}[b]{0.49\textwidth}
         \centering
         \includegraphics[width=\textwidth]{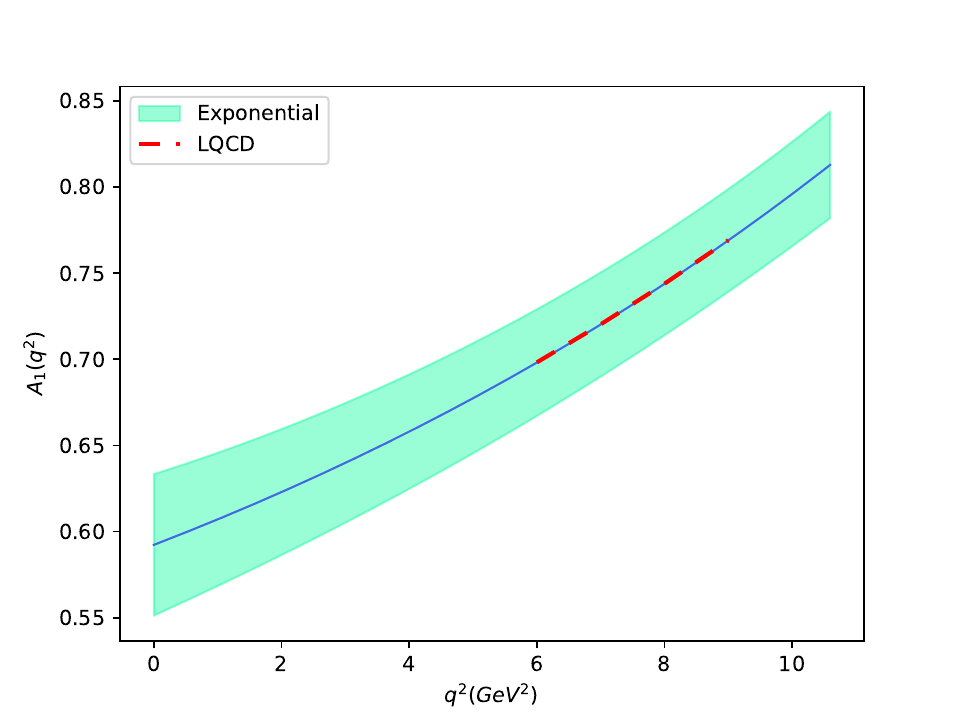}
     \end{subfigure}
     \hfill
     \begin{subfigure}[b]{0.49\textwidth}
         \centering
         \includegraphics[width=\textwidth]{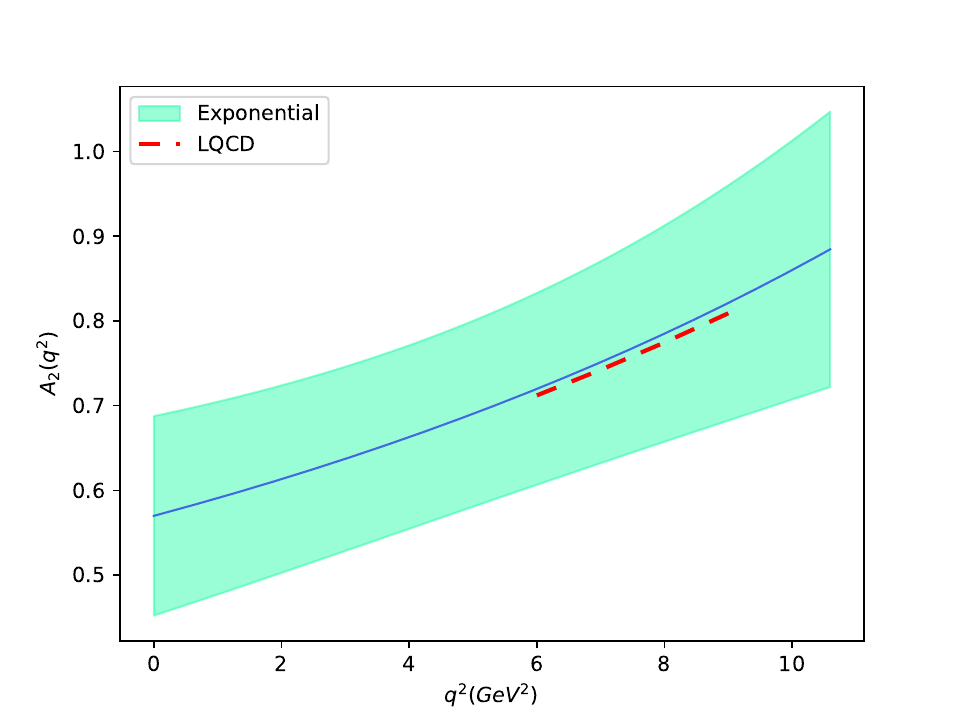}
     \end{subfigure}
     \hfill
          \begin{subfigure}[b]{0.49\textwidth}
         \centering
         \includegraphics[width=\textwidth]{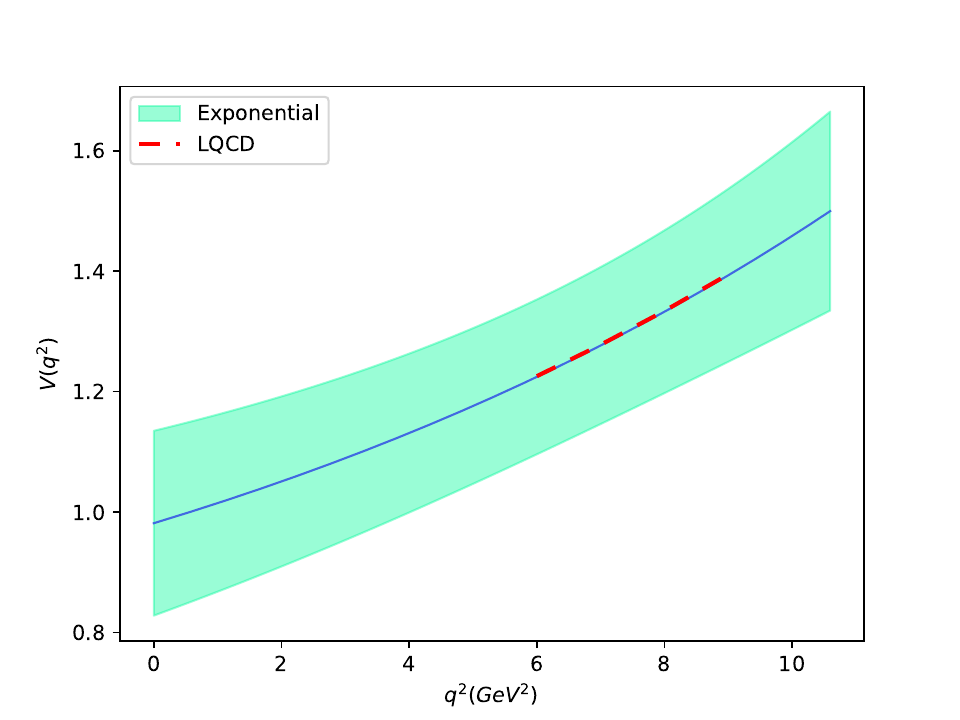}
     \end{subfigure}
     \hfill
             
      \caption{Form factor fits to LQCD data for transition $B_s \to D_s^*$ using exponential formula.} 
      \label{fig: fits BsDS* exp}
\end{figure}

\begin{figure}
     \centering
     \begin{subfigure}[b]{0.49\textwidth}
         \centering
         \includegraphics[width=\textwidth]{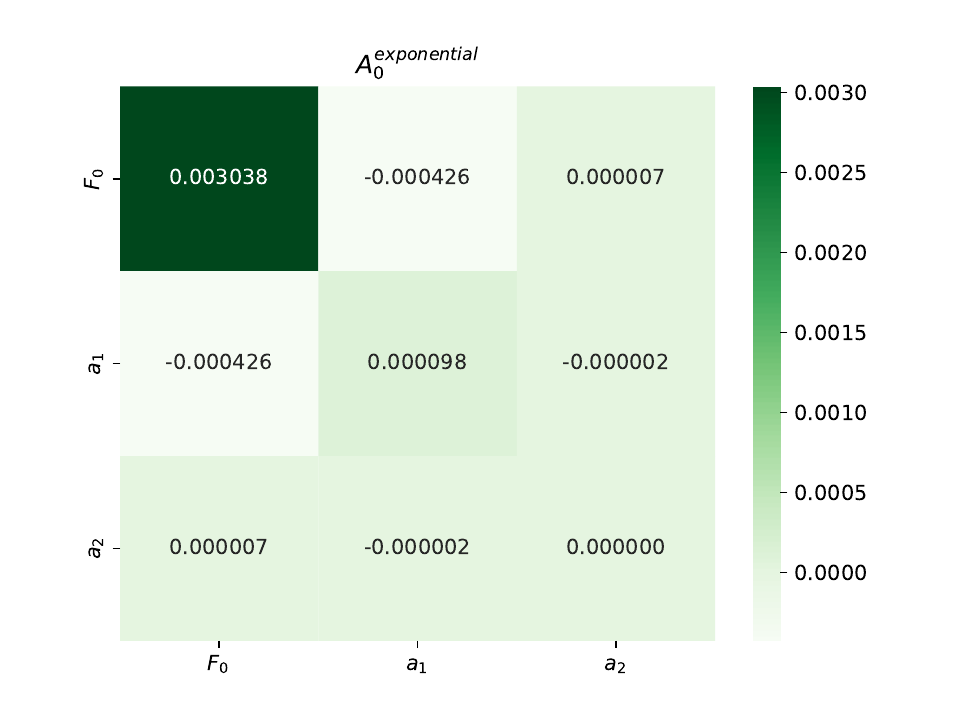}
     \end{subfigure}
     \hfill
     \begin{subfigure}[b]{0.49\textwidth}
         \centering
         \includegraphics[width=\textwidth]{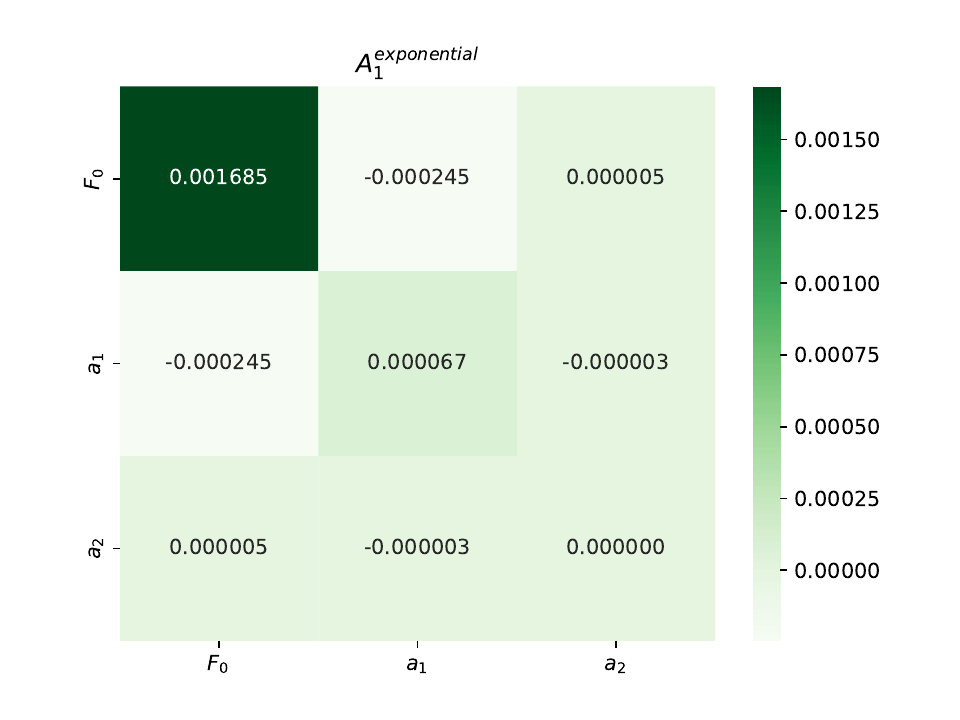}
     \end{subfigure}
     \hfill
     \begin{subfigure}[b]{0.49\textwidth}
         \centering
         \includegraphics[width=\textwidth]{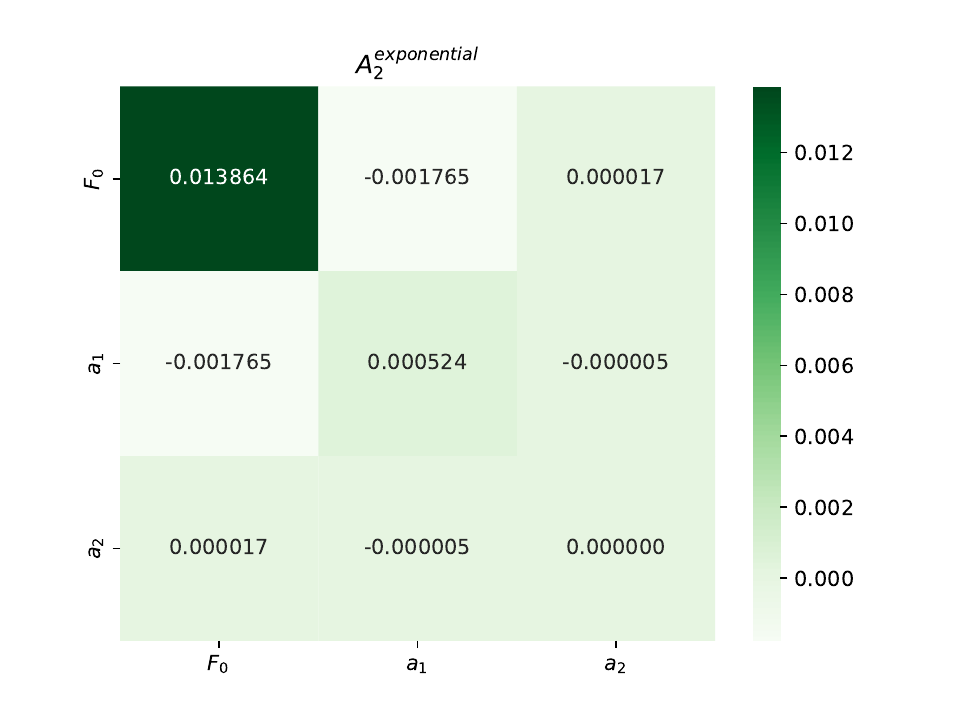}
     \end{subfigure}
     \hfill
          \begin{subfigure}[b]{0.49\textwidth}
         \centering
         \includegraphics[width=\textwidth]{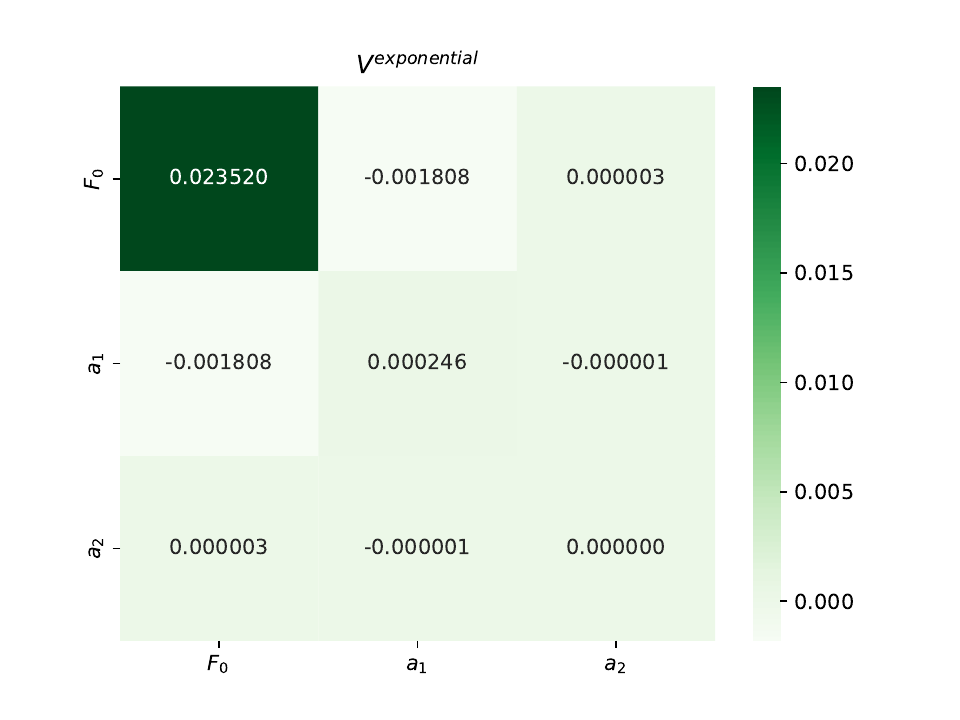}
     \end{subfigure}
     \hfill
             
      \caption{The covariance of parameters in the exponential form factors $V, A_0, A_1, A_2$.} 
      \label{fig: cov-exp-BsDs*}
\end{figure}

\begin{table}[htb]
\caption{Longitudinal, transverse lepton polarization and leptonic convexity parameters for semileptonic transition of $B_s \to D_s \ell \nu_{\ell}$.}
\label{tab: polar_BsDs}
\begin{tabularx}{1\textwidth}{>{\centering\arraybackslash}X | >{\centering\arraybackslash}X | >{\centering\arraybackslash}X |>{\centering\arraybackslash}X | >{\centering\arraybackslash}X | >{\centering\arraybackslash}X| >{\centering\arraybackslash}X| >{\centering\arraybackslash}X| >{\centering\arraybackslash}X| >{\centering\arraybackslash}X }

\hline\hline

 & $\left\langle {P_L^e} \right\rangle $ & $\left\langle {P_L^\mu} \right\rangle  $ & $\left\langle {P_L^\tau} \right\rangle $ & $\left\langle {P_T^e} \right\rangle $ & $\left\langle {P_T^\mu} \right\rangle $ & $\left\langle {P_T^\tau} \right\rangle $ & $\left\langle {C_F^e} \right\rangle $ & $\left\langle {C_F^\mu} \right\rangle $ & $\left\langle {C_F^\tau} \right\rangle $
\\
\hline
dipole & $1.0000$ & $0.9611$ & $-0.3211$  & $-0.0010$  & $-0.1975$ & $-0.8419$ & $-1.5000$  & $-1.4574$  & $-0.2702$
\\
\hline
exp & $1.0000$ & $0.9611$ & $-0.3209$  & $-0.0010$ & $-0.1974$  & $-0.8419$ & $-1.5000$ & $-1.4574$ & $-0.2703$
\\
\hline\hline
\end{tabularx}
\end{table}
For the semileptonic $B_s \to D_s^*$, vector and axial-vector form factors have been computed by HPQCD Collaboration using lattice QCD \cite{Harrison:2021tol} with $z$-expansion parameterization with three parameters. We have used their mean values and correlation matrices to fit with the form factors of $B_s \to D_s^*$, $V(q^2), A_{0,1,2}(q^2)$. In Figs. \ref{fig: fits BsDS* dipole}, \ref{fig: cov-dipole-BsDs*}, \ref{fig: fits BsDS* exp}, \ref{fig: cov-exp-BsDs*}, the results associated to the fitting procedure are plotted. As can be seen from Figs.  \ref{fig: fits BsDS* dipole} and \ref{fig: fits BsDS* exp}, the behavior of the dipole model is more appropriate than exponential, especially in the case of $A_2(q^2)$. Moreover, the covariances between parameters of the dipole are larger than other ones, see Figs. \ref{fig: cov-dipole-BsDs*} and \ref{fig: cov-exp-BsDs*}. In Table \ref{tab: fit_params_BsDs*}, we report the values of $F(0), a_1, a_2$ for the current transition. Using these values in the form factors as well as the helicity amplitudes in Eq. (\ref{Eq: helicity Bs Ds*}), and the differential decay distribution, Eq. (\ref{eq: decay unpol}), one can evaluate the decay rates of $B_s \to D_s^* \ell \nu$ for semi-electronic, semi-muonic, and semi-taunic channels in Table \ref{tab: BsDs*_BR}. For the semi-muonic channel, our values, $(5.2246 \pm 0.4624)\% $ and $(5.2386 \pm 0.4452)\% $ agree well with the PDG data $BR(B_s^0 \to D_s^{* - }{\mu ^ + }{\nu _\mu }) = (5.2 \pm 0.5)\%$ \cite{ParticleDataGroup:2024cfk}. In Ref. \cite{Hassanabadi:2014isa}, the authors obtained $1.35\%$ for the branching fraction of the decay $B_s \to D_s \ell \nu$. Based on the polarized differential decay, the longitudinal and transverse polarizations, the lepton convexity and the longitudinal polarization of a vector meson $D_s^*$ in section \ref{sec: observables} can be obtained, we report the results for the mentioned quantities in addition to the average of forward-backward asymmetries in Tables \ref{tab: FB_BsDs*}, \ref{tab: polar_BsDs*}. The observable $F_L^\tau $ for vector meson $D_s^*$ were reported 0.433 and 0.43 in Refs \cite{Das:2019cpt, Hu:2019bdf} respectively. Our values for longitudinal polarization of $D_s^*$ in Table \ref{tab: FB_BsDs*}, 0.4401 and 0.4391, are close to them.
We apply our results to find the ratio 
\begin{equation}
    \frac{{\Gamma (B_s^0 \to D_s^ - {\mu ^ + }{\nu _\mu })}}{{\Gamma (B_s^0 \to D_s^{* - }{\mu ^ + }{\nu _\mu })}} = 0.4415 \pm 0.1860,
\end{equation}
with the dipole parametrization and 
\begin{equation}
    \frac{{\Gamma (B_s^0 \to D_s^ - {\mu ^ + }{\nu _\mu })}}{{\Gamma (B_s^0 \to D_s^{* - }{\mu ^ + }{\nu _\mu })}} = 0.4406 \pm 0.1854,
\end{equation}
with the exponential one. Those are acceptable when compared with the reported measurement value of $0.464(45)$ by LHCb collaboration \cite {LHCb:2020cyw}. Moreover, another important quantity is the ratio of $R(D_s)$, using in the lepton
universality with an interesting picture emerging, which is calculated as
\begin{equation}
    R({D_s}) = \frac{{BR(B_s^0 \to D_s^ - {\tau ^ + }{\nu _\tau })}}{{BR(B_s^0 \to D_s^ - {\mu ^ + }{\nu _\mu })}} = 0.2983 \pm 0.1229
\end{equation}
with the dipole formula and
\begin{equation}
    R({D_s}) = \frac{{BR(B_s^0 \to D_s^ - {\tau ^ + }{\nu _\tau })}}{{BR(B_s^0 \to D_s^ - {\mu ^ + }{\nu _\mu })}} = 0.2984 \pm 0.1230
\end{equation}
with the exponential formula. Zhang \textit{et al.}, reported $R({D_s}) = 0.334 \pm 0.017$ for this quantity and also found the form factors of $B_s \to D_s$ at the largest recoil point, $f_{ + ,0}^{{B_s} \to {D_s}}(0) = 0.533_{ - 0.094}^{ + 0.112}$ \cite{Zhang:2022opp}. Ours, $f_{ + ,0}^{{B_s} \to {D_s}} (0) = 0.67 \pm 0.01$, is in consistent with them. The ratio $R(D_s) = 0.320_{ - 0.009}^{ + 0.009}$ was also reported in Ref. \cite{Zhou:2019stx}.
Another ratio which may measured by various groups, is $R(D_s^*)$. We obtain that in both models such that
\begin{equation}
    R(D_s^*) = 0.2503 \pm 0.0232 \text{ (dipole), } R(D_s^*) = 0.2491 \pm 0.0223 \text{ (exponential)}. 
\end{equation}

This quantity was found to be $R(D_s^*) = 0.2490(60)(35)$ within the Lattice QCD formalism
\cite{Harrison:2021tol}. Moreover, Zhou \textit{et al.}, obtained $R(D_s^*) = 0.251_{ - 0.003}^{ + 0.002}$ using Bethe–Salpeter method \cite{Zhou:2019stx}.
\par
\begin{figure}[h]
    \includegraphics[width=0.5\textwidth]{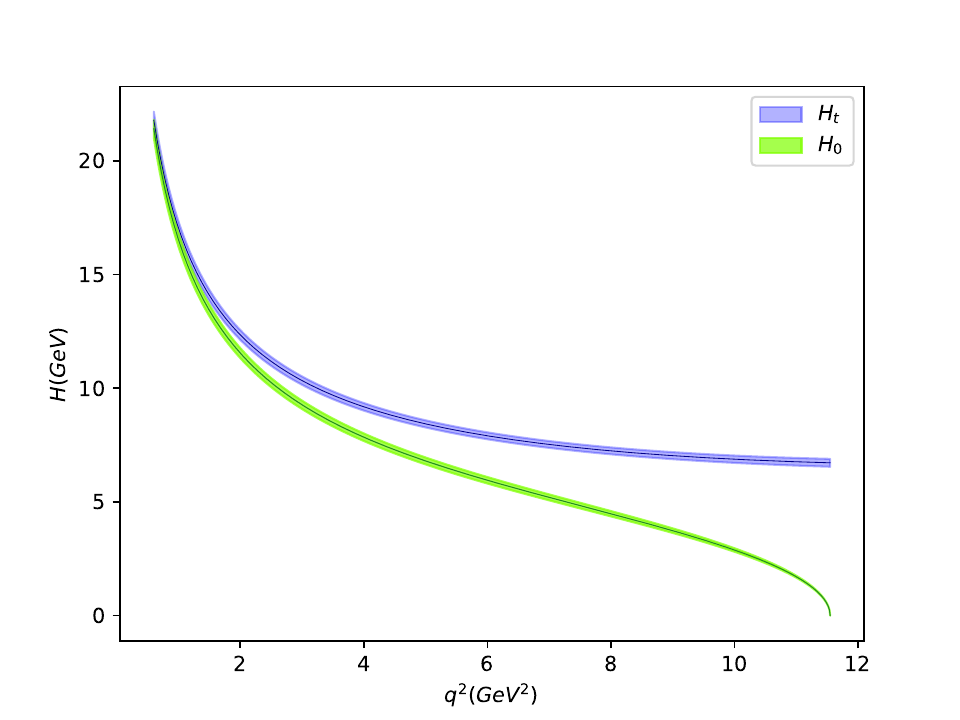}
    \caption{The helicity amplitudes versus $q^2$ for $B_s \to D_s$ decay.}
    \label{fig: helicity BsDs}
\end{figure}
\begin{figure}[h]
    \includegraphics[width=0.5\textwidth]{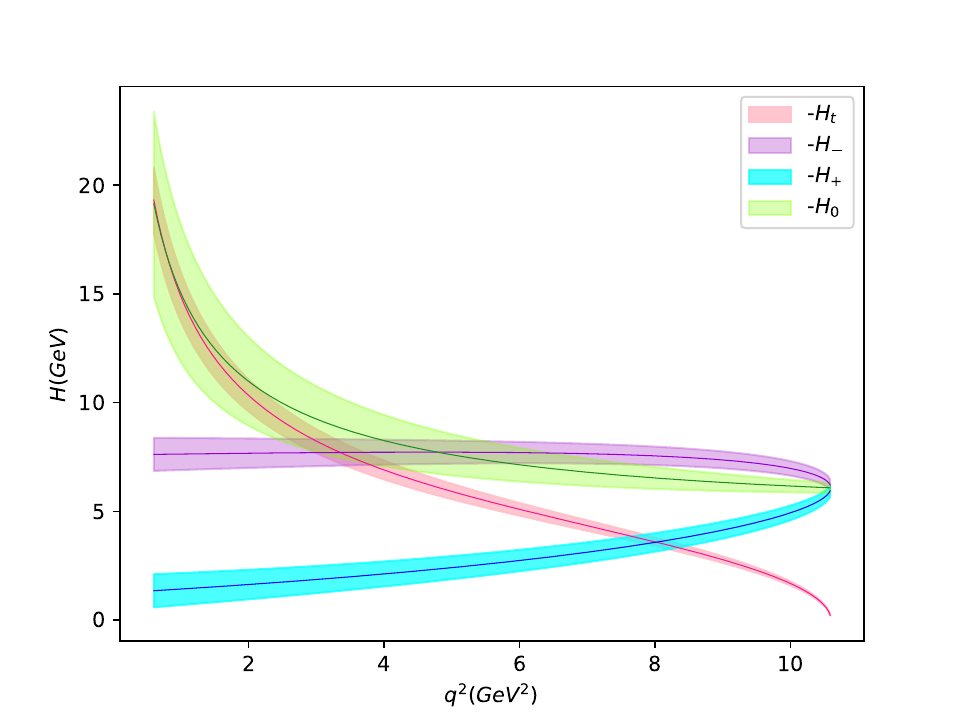}
    \caption{The helicity amplitudes in the transition of $B_s \to D_s^*$ decay.}
    \label{fig: helicity BsDs*}
\end{figure}
\begin{figure}[h]
    \includegraphics[width=0.5\textwidth]{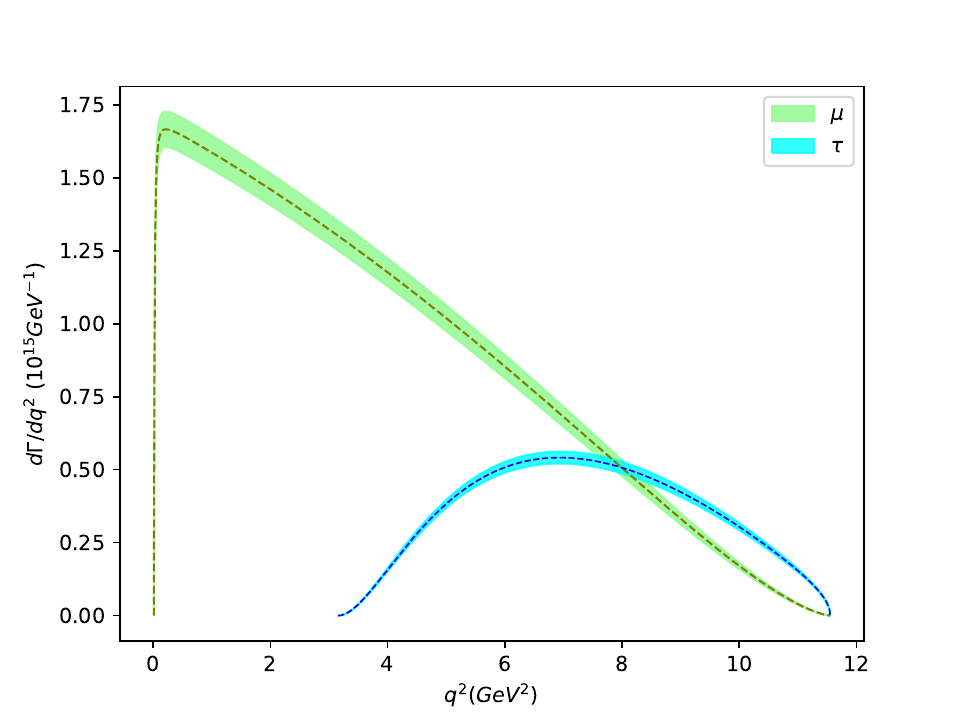}
    \caption{The differential decay width for the semi-muonic and semi-taunic channels of $B_s \to D_s$.}
    \label{fig: DW_BsDs}
\end{figure}
\begin{figure}[h]
    \includegraphics[width=0.5\textwidth]{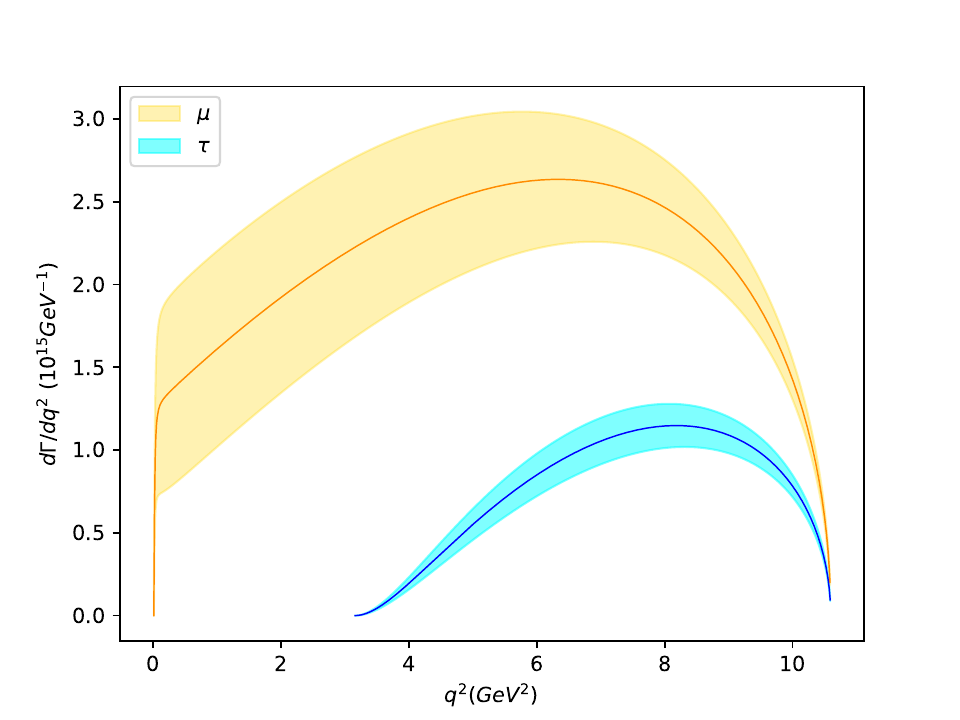}
    \caption{Same as Fig. \ref{fig: DW_BsDs} for $B_s \to D_s^*$.}
    \label{fig: DW_BsDs*}
\end{figure}
\begin{figure}
     \centering
     \begin{subfigure}[b]{0.49\textwidth}
         \centering
         \includegraphics[width=\textwidth]{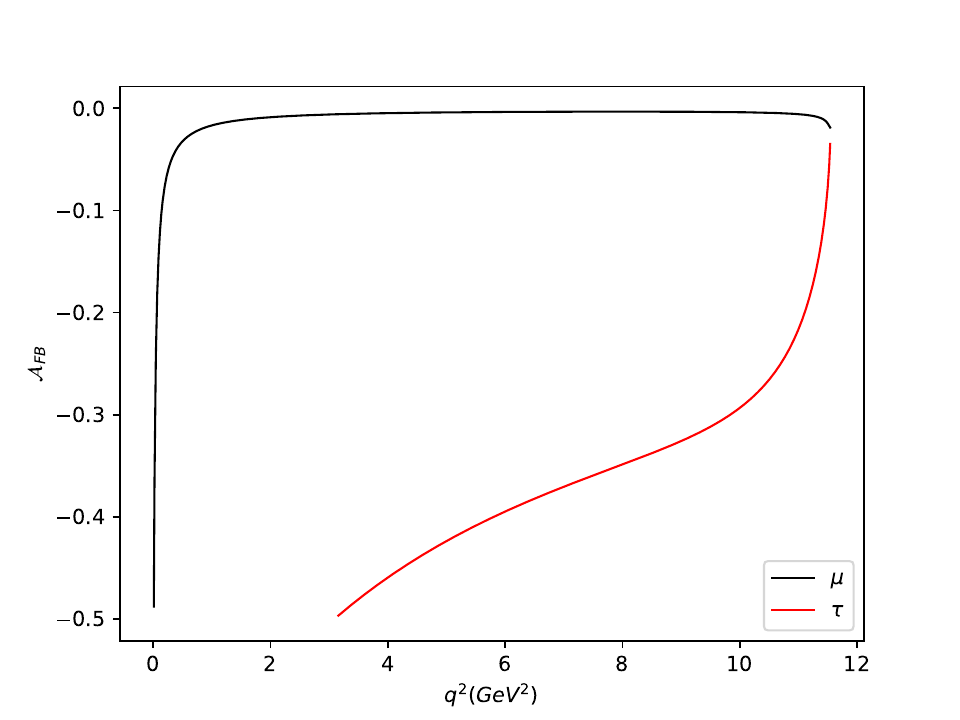}
     \end{subfigure}
     \hfill
     \begin{subfigure}[b]{0.49\textwidth}
         \centering
         \includegraphics[width=\textwidth]{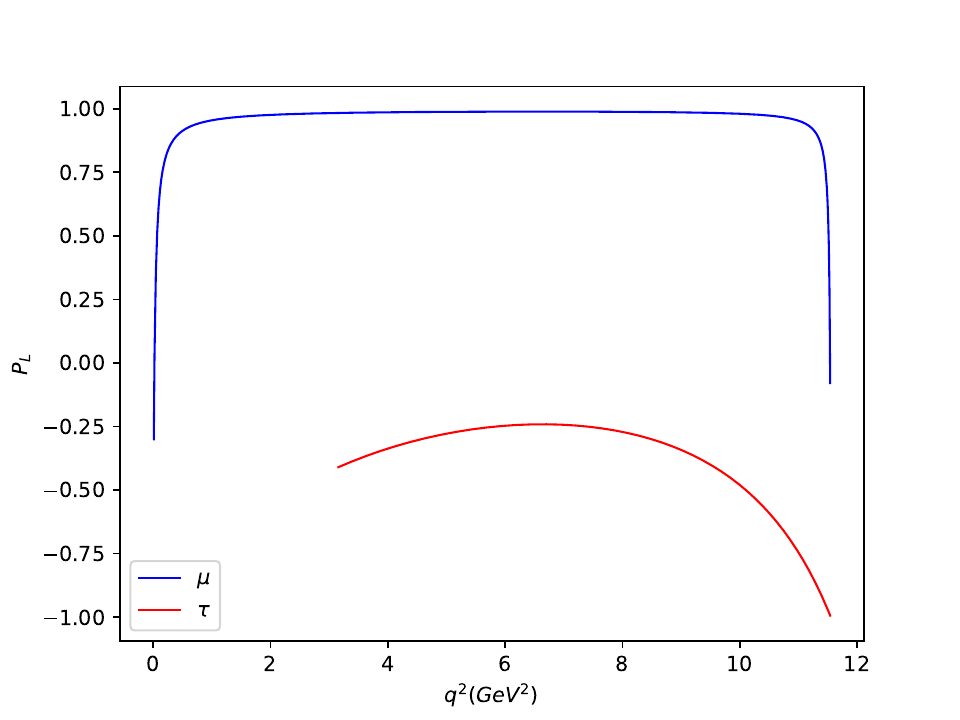}
     \end{subfigure}
     \hfill
     \begin{subfigure}[b]{0.49\textwidth}
         \centering
         \includegraphics[width=\textwidth]{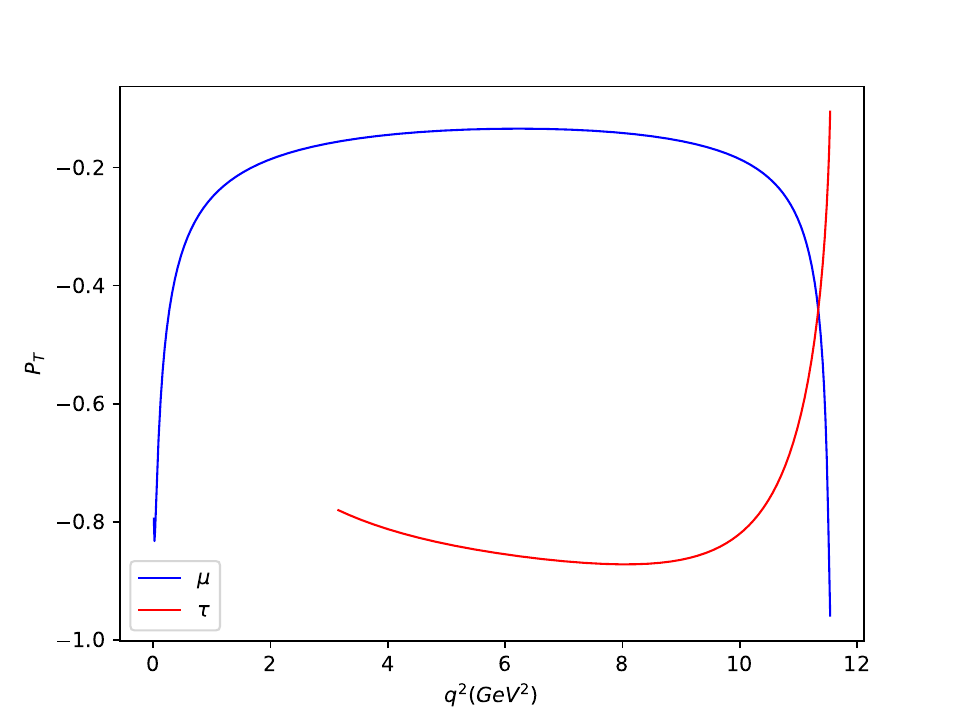}
     \end{subfigure}
     \hfill             
      \caption{The forward-backward asymmetries, longitudinal and transverse polarization of a charged lepton of the decays $B_s \to D_s \ell^+ \nu$. } 
      \label{fig: FB polar BsDs}
\end{figure}
\begin{figure}
     \centering
     \begin{subfigure}[b]{0.49\textwidth}
         \centering
         \includegraphics[width=\textwidth]{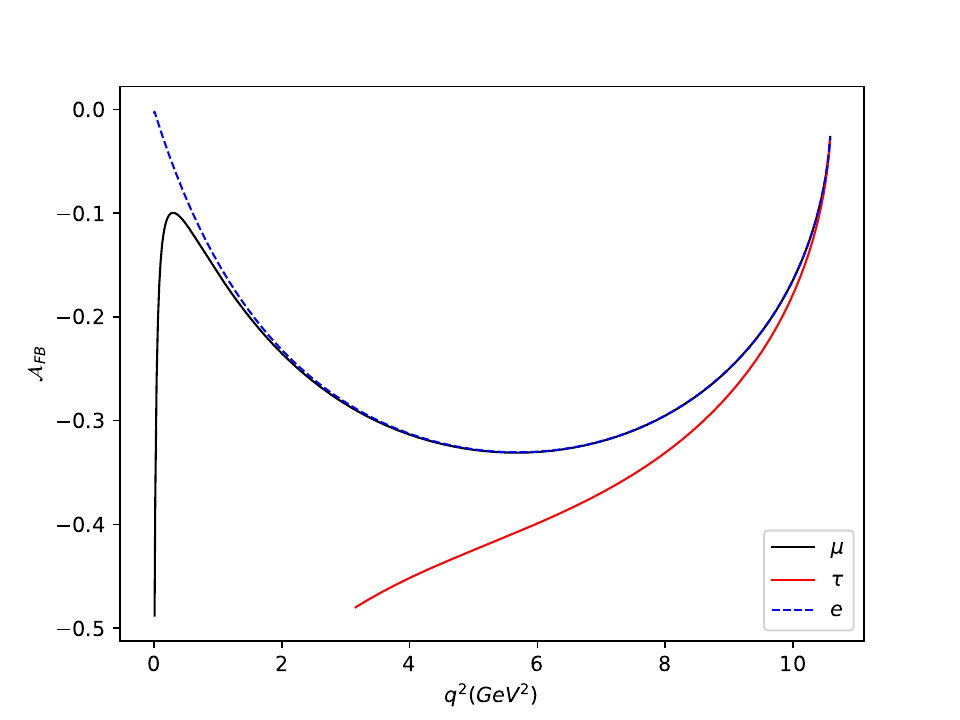}
     \end{subfigure}
     \hfill
     \begin{subfigure}[b]{0.49\textwidth}
         \centering
         \includegraphics[width=\textwidth]{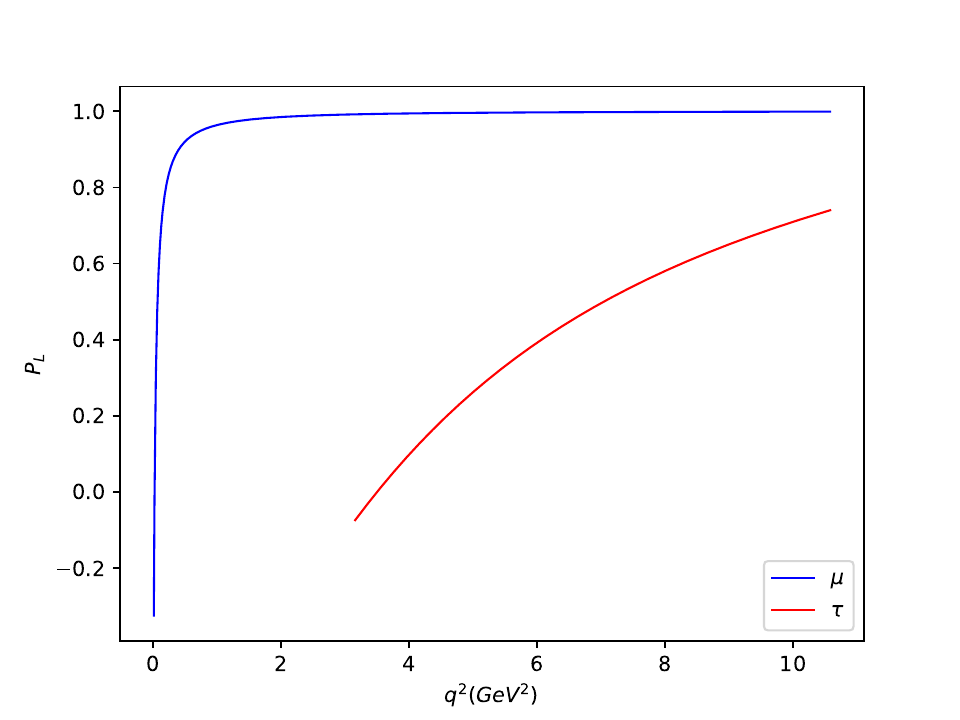}
     \end{subfigure}
     \hfill
     \begin{subfigure}[b]{0.49\textwidth}
         \centering
         \includegraphics[width=\textwidth]{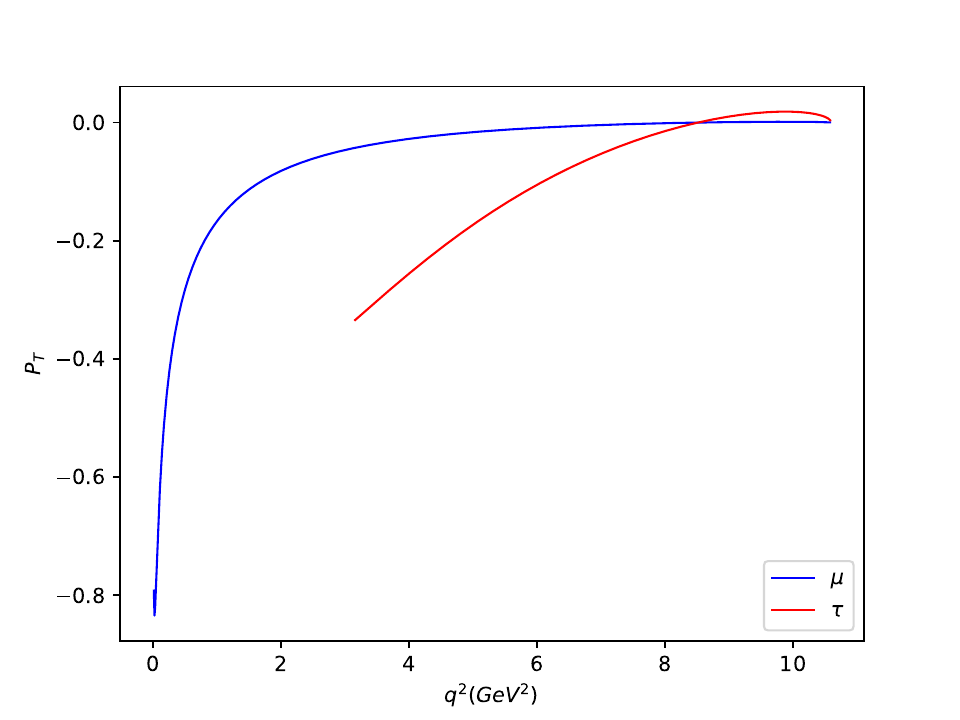}
     \end{subfigure}
     \hfill             
      \caption{Same as Fig. \ref{fig: FB polar BsDs} for the decay $B_s \to D_s^* \ell \nu$} 
      \label{fig: FB polar BsDs*}
\end{figure}

Using our form factors in the dipole case, we can construct helicity amplitudes with their error bands. They are plotted in Figs. \ref{fig: helicity BsDs}, \ref{fig: helicity BsDs*} for $B_s \to D_s$ and $B_s \to D_s^*$ respectively. Our helicity amplitudes for $B_s \to D_s^*$ are equivalent to those in Ref. \cite{Harrison:2021tol} times a minus sign. At large value of $q^2$, $H_0, H_{\pm}$ will be equal. Also, $H_0$ and $H_t$ have a singularity at $q^2 = 0$ as one would expect from their relations for both transitions $B_s \to D_s$ and $B_s \to D_s^*$. The differential decay rates for $B_s \to D_s \mu \nu$ and $B_s \to D_s \tau \nu$ are compared in Fig. \ref{fig: DW_BsDs}. The dashed lines are associated with the central values and the blue and green bands represent their errors. Fig. \ref{fig: DW_BsDs*} show the behaviour of $\frac{d\Gamma}{dq^2}$ for ${B_s^0 \to D_s^{* - }{\ell ^ + }{\nu _\ell }}$, where $\ell = \mu, \tau$ as a function of $q^2$. The $q^2$ dependence of the forward-backward asymmetries, the leptonic longitudinal and transverse polarizations are shown in Figs. \ref{fig: FB polar BsDs} and \ref{fig: FB polar BsDs*} for both transitions. At the zero recoil point ($q^2 = q^2_{max}$), $\mathcal{A}_{FB}$ and $P_T$ in Fig. \ref{fig: FB polar BsDs*}, will be zero as one would expect from their relations.

\begin{table}[htb]
\caption{Fitted parameters $F(0), a_1$, and $a_2$ for $V$, $A_0$, $A_1$ and $A_2$ in transition of $B_s \to D_s^*$.}
\label{tab: fit_params_BsDs*}
\begin{tabularx}{1\textwidth}{>{\centering\arraybackslash}X | >{\centering\arraybackslash}X | >{\centering\arraybackslash}X |>{\centering\arraybackslash}X }

\hline\hline
      & $F(0)$ & $a_1$ & $a_2$
      \\
\hline
 $V^{dipole}({q^2})$ & $0.9764 \pm 0.1520$ & $1.4011 \pm 0.6031$ & $0.2931 \pm 0.8256$
\\
\hline
 $V^{exponential}({q^2})$ & $0.9817 \pm 0.1534$ & $0.0327 \pm 0.0157 $ & $0.0007 \pm 0.0003$
\\
\hline
 $A_0^{dipole}({q^2})$ & $0.6195 \pm 0.0495$ & $2.0026 \pm 0.3044$  &  $1.1041 \pm 0.5948$
\\
\hline
 $A_0^{exponential}({q^2})$ & $0.6146 \pm 0.0551$ & $0.0452 \pm 0.0099$ & $0.0005 \pm 0.0003$
\\
\hline
 $A_1^{dipole}({q^2})$ & $0.5905 \pm 0.0404$ & $1.1606 \pm 0.3248$ & $-0.0653 \pm 0.7595$
 \\
\hline
 $A_1^{exponential}({q^2})$ & $0.5923 \pm 0.0410$ & $0.0241 \pm 0.0082$ & $0.0005 \pm 0.0004$
\\
\hline
 $A_2^{dipole}({q^2})$ & $0.5819 \pm 0.0918 $ & $1.4000 \pm 0.4201$ & $0.0050 \pm 0.0027$
\\
\hline
$A_2^{exponential}({q^2})$ & $0.5699 \pm 0.1177$ & $0.0355 \pm 0.0229$ & $0.0006 \pm 0.0004$
\\
\hline\hline
\end{tabularx}
\end{table}

\begin{table}[htb]
\caption{Same as Table \ref{tab: BsDs_BR} for semileptonic transition of $B_s \to D_s^* \ell \nu_{\ell}$.}
\label{tab: BsDs*_BR}
\begin{tabularx}{1\textwidth}{>{\centering\arraybackslash}X | >{\centering\arraybackslash}X | >{\centering\arraybackslash}X |>{\centering\arraybackslash}X | >{\centering\arraybackslash}X}

\hline\hline

Mode & $\Gamma$ with dipole & $\Gamma$ with exponential & $BR$ with dipole  & $BR$ with exponential
\\
\hline
$B_s^0 \to D_s^{*-} {\mu ^ + }{\nu _\mu }$ & $22.5097 \pm 1.9921$ & $22.5702 \pm 1.9182$ & $5.2246 \pm 0.4624$  & $5.2386 \pm 0.4452$
\\
\hline
$B_s^0 \to D_s^{*-} {e^ + }{\nu _e}$ & $22.6055 \pm 0.5942$ & $22.6720 \pm 0.6018$ & $5.2468 \pm 0.1379$ & $5.2623 \pm 0.1397$
\\
\hline
$B_s^0 \to D_s^ {*-} {\tau ^ + }{\nu _\tau }$ & $5.6349 \pm 0.1549$ & $5.6228 \pm 0.1549$ & $1.3079 \pm 0.0359$ & $1.3051 \pm 0.0360$
\\
\hline\hline
\end{tabularx}
\end{table}

\begin{table}[htb]
\caption{Forward-backward asymmetry and longitudinal polarization of vector meson for semileptonic transition of $B_s \to D_s^* \ell \nu_{\ell}$.}
\label{tab: FB_BsDs*}
\begin{tabularx}{1\textwidth}{>{\centering\arraybackslash}X | >{\centering\arraybackslash}X | >{\centering\arraybackslash}X |>{\centering\arraybackslash}X | >{\centering\arraybackslash}X | >{\centering\arraybackslash}X | >{\centering\arraybackslash}X  }

\hline\hline

 & $\left\langle {\mathcal{ A} _{FB}^e} \right\rangle $ & $\left\langle {\mathcal{ A} _{FB}^\mu} \right\rangle $ & $\left\langle {\mathcal{ A} _{FB}^\tau} \right\rangle $ & $\left\langle {F_L^e} \right\rangle $& $\left\langle {F_L^\mu } \right\rangle $ & $\left\langle {F_L^\tau } \right\rangle $
\\
\hline
dipole & $-0.2700$ & $-0.2743$ & $-0.3295$ & $0.4983$& $0.4981$& $0.4401$
\\
\hline
exponential & $-0.2692$ & $-0.2736$ & $-0.3295$ &$0.4997$ & $0.4994$& $0.4391$
\\
\hline\hline
\end{tabularx}
\end{table}

\begin{table}[htb]
\caption{Same as Table \ref{tab: polar_BsDs} for semileptonic transition of $B_s \to D_s^* \ell \nu_{\ell}$.}
\label{tab: polar_BsDs*}
\begin{tabularx}{1\textwidth}{>{\centering\arraybackslash}X | >{\centering\arraybackslash}X | >{\centering\arraybackslash}X |>{\centering\arraybackslash}X | >{\centering\arraybackslash}X | >{\centering\arraybackslash}X| >{\centering\arraybackslash}X| >{\centering\arraybackslash}X| >{\centering\arraybackslash}X| >{\centering\arraybackslash}X }

\hline\hline

 & $\left\langle {P_L^e} \right\rangle $ & $\left\langle {P_L^\mu} \right\rangle  $ & $\left\langle {P_L^\tau} \right\rangle $ & $\left\langle {P_T^e} \right\rangle $ & $\left\langle {P_T^\mu} \right\rangle $ & $\left\langle {P_T^\tau} \right\rangle $ & $\left\langle {C_F^e} \right\rangle $ & $\left\langle {C_F^\mu} \right\rangle $ & $\left\langle {C_F^\tau} \right\rangle $
\\
\hline
dipole & $1.0000$ & $0.9865$ & $0.5198$  & $-0.0002$  & $-0.0432$ & $-0.0499$ & $-0.3712$  & $-0.3582$  & $-0.0496$
\\
\hline
exp & $1.0000$ & $0.9866$ & $0.5201$  & $-0.0002$  & $-0.0432$ & $-0.0491$ & $-0.3744$  & $-0.3611$  & $-0.0486$
\\
\hline\hline
\end{tabularx}
\end{table}

Furthermore, in Table \ref{tab: nonleptonic}, we give the predictions for the branching ratios of two body nonleptonic decays of $B_s$ to a $D_s$ meson along with a pseudoscalar or vector in their final states. The results are compared with PDG \cite{ParticleDataGroup:2024cfk}, factorization approach \cite{Azizi:2008ty}, and quark models \cite{Faustov:2012mt, Albertus:2014eqa}. 
\begin{table}[htb]
\caption{Branching fractions for the indicated decay channels ${B_s} \to PP,{B_s} \to PV$ in percentage}
\label{tab: nonleptonic}
\begin{tabularx}{1\textwidth}{>{\centering\arraybackslash}X | >{\centering\arraybackslash}X | >{\centering\arraybackslash}X |>{\centering\arraybackslash}X | >{\centering\arraybackslash}X | >{\centering\arraybackslash}X| >{\centering\arraybackslash}X  }

\hline\hline

Modes & Exponential & Dipole & PDG \cite{ParticleDataGroup:2024cfk} & \cite{Azizi:2008ty} & \cite{Faustov:2012mt} & \cite{Albertus:2014eqa}
\\
\hline
${B_s} \to D_s^ - {\pi ^ + }$ & $0.3197 \pm 0.0118$ & $0.3197 \pm 0.0111$ & $(2.98 \pm 0.14) \times {10^{ - 1}}$ & - & $0.35$ & $0.53$
\\
\hline
${B_s} \to D_s^ - {K^ + }$& $0.0242 \pm 0.0009$  & $0.0242 \pm  0.0008 $ & $(2.25 \pm 0.12) \times {10^{ - 2}}$ & -& $0.028$& $0.04$
\\
\hline
${B_s} \to D_s^ - D_s^ + $& $1.1221 \pm 0.0341$  & $1.1227 \pm 0.0332$ & $(4.4 \pm 0.5) \times {10^{ - 1}}$ & -& $1.1$ & -
\\
\hline
${B_s} \to D_s^ - {\rho ^ + }$ & $0.0165 \pm 0.0006$ & $0.0165 \pm 0.0006$ & $(6.8 \pm 1.4) \times {10^{ - 1}}$ &- & $0.94$ & $1.26$
\\

\hline
${B_s} \to D_s^ - D_s^{* + }$& $0.1890 \pm 0.0080$ & $0.1889 \pm 0.0079$ & - & $(2.62 \pm 0.93) \times {10^{ - 1}}$ & $1.0$& -
\\
\hline\hline
\end{tabularx}
\end{table}

\section{CONCLUSIONS}
\label{section: conclusion}
We have provided a detailed helicity analysis of semileptonic decays $B_s \to D_s^{(*)} \ell \nu$. Two parameterizations, dipole and exponential, have been considered for the form factors of these weak decays. We have performed fitting to the LQCD data \cite{Harrison:2021tol} and \cite{McLean:2019qcx}, and obtained free parameters as well as their covariances. 
We have applied the form factors to obtain the two body nonleptonic decays $B_s \to D_s + P(V)$. The $q^2$ dependence of decay widths and physical observables have been plotted. The branching fractions, forward-backward asymmetry, longitudinal and transverse polarizations of the charged lepton as well as the vector meson have been calculated. Our obtained branching fractions, especially in the semi-muonic channels, are in good agreement with the PDG data \cite{ParticleDataGroup:2024cfk}.
\par
\textbf{Acknowlegements}
\par
I would like to thank De-Liang Yao for useful discussions. This work is supported by the
Fundamental Research Funds for the Central Universities under Contract No. 531118010379.

\end{document}